\documentclass[acmsmall]{acmart}

\AtBeginDocument{%
  \providecommand\BibTeX{{%
    \normalfont B\kern-0.5em{\scshape i\kern-0.25em b}\kern-0.8em\TeX}}}

\setcopyright{acmcopyright}
\acmJournal{TOIS}
\acmYear{2022} \acmVolume{x} \acmNumber{x} \acmArticle{1} \acmMonth{1} \acmPrice{15.00}\acmDOI{10.xxx}

\newcommand{\ie}{\emph{i.e., }}
\newcommand{\eg}{\emph{e.g., }}

\newcommand{\etc}{\emph{etc.}}
\newcommand{\wrt}{\emph{w.r.t. }}
\newcommand{\cf}{\emph{cf. }}

\newcommand{\zy}[1]{{#1}}
\newcommand{\zytoisnew}[1]{{#1}}
\newcommand{\zytois}[1]{{#1}}

\usepackage{enumitem}
\usepackage{verbatim}
\usepackage[ruled]{algorithm2e}
\usepackage{multirow}
\usepackage{subfigure} 
\usepackage{ragged2e}
\usepackage{caption}
\usepackage{graphicx}
\usepackage[normalem]{ulem}
\useunder{\uline}{\ul}{}
\usepackage{amsmath,bm}
\usepackage{setspace}
\usepackage{booktabs}
\begin{document}

\title{Addressing Confounding Feature Issue for Causal Recommendation}


\author{Xiangnan He}
\affiliation{%
  \institution{University of Science and Technology of China}
  \city{Hefei}
  \country{China}
  \postcode{230027}}
\email{xiangnanhe@gmail.com}

\author{Yang Zhang}
\authornote{Corresponding author.}
\authornote{Partial work done at Tencent.}
\email{zy2015@mail.ustc.edu.cn}
\affiliation{%
  \institution{University of Science and Technology of China}
  \city{Hefei}
  \country{China}
  \postcode{230027}
}

\author{Fuli Feng}
\affiliation{%
  \institution{University of Science and Technology of China}
  \city{Hefei}
  \country{China}
  \postcode{230027}
}
\email{fulifeng93@gmail.com}

\author{Chonggang Song}
\affiliation{%
 \institution{WeChat, Tencent}
 \city{Shenzhen}
 \country{China}
 \postcode{518054}}
 \email{jerrycgsong@tencent.com}

\author{Lingling Yi}
\affiliation{%
  \institution{WeChat, Tencent}
  \city{Shenzhen}
  \country{China}
  \postcode{518054}}
  \email{chrisyi@tencent.com}

\author{Guohui Ling}
\affiliation{%
  \institution{WeChat, Tencent}
  \city{Shenzhen}
  \country{China}
  \postcode{518054}
}
\email{randyling@tencent.com}

\author{Yongdong Zhang}
\email{zhyd73@ustc.edu.cn}
\affiliation{%
  \institution{University of Science and Technology of China}
  \city{Hefei}
  \country{China}
  \postcode{230027}}

\renewcommand{\shortauthors}{Xiangnan He, et al.}

\begin{abstract}
 In recommender system, some feature directly affects whether an interaction would happen, making the happened interactions not necessarily indicate user preference. 
For instance, short videos are objectively easier to be finished even though the user does not like the video. We term such feature as \textit{confounding feature}, and video length is a confounding feature in video recommendation.  
If we fit a model on such interaction data, just as done by most data-driven recommender systems, the model will be biased to recommend short videos more, and deviate from user actual requirement. 

This work formulates and addresses the problem from the causal perspective. 
Assuming there are some factors affecting both the confounding feature and other item features, \eg the video creator, we find the confounding feature opens a backdoor path behind user-item matching and introduces spurious correlation. 
To remove the effect of backdoor path, we propose \zy{a framework named \textit{Deconfounding Causal Recommendation} (DCR)}, which performs intervened inference with \textit{do-calculus}. Nevertheless, evaluating \textit{do-calculus} requires to sum over the prediction on all possible values of confounding feature, significantly increasing the time cost. 
To address the efficiency challenge, we further propose a mixture-of-experts (MoE) model architecture, modeling each value of confounding feature with a separate expert module. Through this way, we retain the model expressiveness with few additional costs. 
We demonstrate DCR on the backbone model of neural factorization machine (NFM), showing that DCR leads to more accurate prediction of user preference with small inference time cost. We release our code at: https://github.com/zyang1580/DCR.
\end{abstract}

\maketitle

\section{Introduction}

Most recommendation methods assume that the interactions are caused (or captured) by the matching between user reference and item features~\cite{deep-matching,e-commerce,deepFM}. However, some item feature could directly affect the happening of an interaction in practice. For instance, videos with short length are easier to be finished, and news articles with attractive title or cover image are easier to be clicked, even though the user does not like the content actually. We name such feature as \textit{confounding feature}, which results in the happened interactions not faithfully reflect user preference. If fitting recommender model on such interaction data, the model will learn the shortcut of the confounding feature, \eg assigning higher scores for short videos. The biased recommendation is undesired, and even worse, makes the recommender system vulnerable to attack --- e.g., the video creator may purposefully upload short videos to make them easier to be recommended. 

How to avoid the influence of the confounding feature? An intuitive solution is removing it from the input features, \eg fitting CTR model on other item features. However, since the interaction data is generated partially due to the confounding feature, the matching model is very likely to learn the effect implicitly (\eg the item embeddings may encode the semantics of confounding feature). 
Another solution is training the model with the confounding feature, while removing it during inference to eliminate its effect in the ranking function\footnote{This solution is originally designed for addressing some existing fairness or biases issues in recommendation~\cite{clickbait,MACR} instead of the proposed confounding feature issue, but it can be utilized to deal with the proposed issue, intuitively.}.
The effectiveness of this solution depends on the quality of disentanglement --- how well we can separate the effect of confounding feature and other features~\cite{MACR}. Nevertheless, disentangled learning remains to be an open problem which limits the efficacy of this solution.

\begin{figure}
    \centering
    \includegraphics[width=0.8\textwidth]{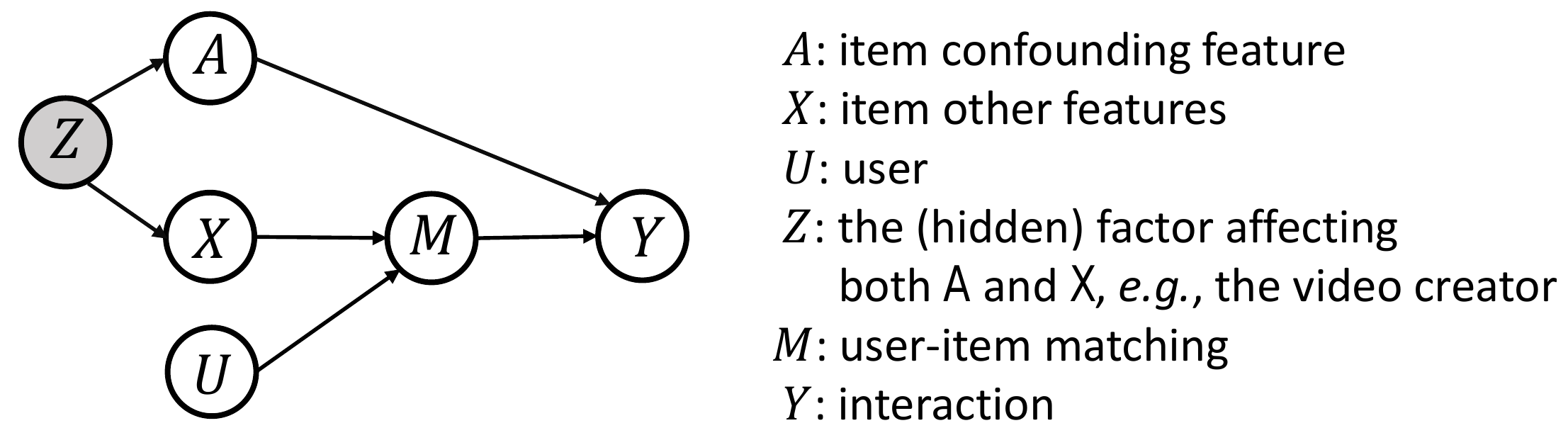}
    \caption{Causal graph to describe the generation process of interactions. $X\longleftarrow Z \longrightarrow A \longrightarrow Y$ is the backdoor path that brings spurious correlations between $X$ and $Y$. }
    \label{fig:our_graph_z}
\end{figure}


To explore the root reason of how confounding feature affects, we abstract the interaction generation process as a casual graph, in Figure~\ref{fig:our_graph_z}.
Let $A$ and $X$ denote the confounding feature and other item features, which affect the interaction (node $Y$) directly ($A \longrightarrow Y$) and through matching with user preference ($\{U,X\} \longrightarrow M \longrightarrow Y$), respectively.
Since $A$ and $X$ describe the same item, they are inevitably affected by some (hidden) factor $Z$, \eg the intention of the video creator.
When learning recommender models on the interaction data, clearly, the effect of $A$ will be counted in the model prediction. More profoundly, \zy{the confounding feature }$A$ opens the backdoor path $X \longleftarrow Z \longrightarrow A \longrightarrow Y$, bringing some spurious correlations between $X$ and $Y$.    
To make the recommendation free from the impact of A, we need to estimate the causal effect of X (or equivalently, M) on Y. To this end, we need to cut off the backdoor path 
through intervention~\cite{pearl2009causality}.
However, it is hard to conduct interventional experiments on X since item features are usually static and unchangeable. 
An alternative way is \textit{do-calculus}~\cite{pearl2009causality}, which can achieve the same effect of intervention on observed data. 
Particularly, we propose a Deconfounding Causal Recommendation (DCR) framework that approaches user-item matching as $P(Y| U, do(X))$. During training, we estimate the correlation $P(Y|U, X, A)$ to fit the historical interaction data, since it caters to the generation of interactions. 
Whereas during inference, we use the intervened $P(Y| U, do(X))$ as the ranking function. 
Taking one step further, according to  the causal graph in Figure~\ref{fig:our_graph_z} \zy{and \textit{the backdoor adjustment}~\cite{pearl2009causality}}, $P(Y|U, do(X))$ is equal to $\sum_{a\in\mathcal{A} } P(Y|U,X,a)P(a)$. Which means, we need to iterate over all values of the confounding feature, then conduct a weighted sum on $P(Y|U, X, a)$. This significantly increases the time cost in the inference stage --- by $|\mathcal{A}|$ times\footnote{$\mathcal{A}$ denotes \zy{all possible values of the confounding feature, \ie the sample space of $A$. $|\mathcal{A}|$ denotes the size of $\mathcal{A}$.}} --- making the solution prohibitive for practical use.
To address this challenge, we propose a mixture-of-experts (MoE) model architecture for our DCR framework. Specifically, we use a shared backbone model to capture the matching between $U$ and $X$, the output of which is fed into a separated expert module designed for each value of the confounding feature. Since the $U$-$X$ matching part is shared for all experts, the time cost of evaluating $P(Y|U, do(X))$ can be reduced largely. Meanwhile, 
modeling each confounding feature value with a separated expert
ensures the modeling fidelity, compared with the approximation method used by \cite{wenjiekdd,PDA}\footnote{The widely used approximation method is: $P(Y|U, do(X))\approx P(Y|U,X, \sum_{a \in \mathcal{A}} a P(a))$, which however uses the same parameterization for all confounding feature values. }.

The main contributions of our work are summarized as follows:
\begin{itemize}
     \item We study a new problem of \textit{confounding feature} in recommender system, analyzing its damage effect from the causal perspective.  
    \item We propose a new solution framework DCR to address the problem with intervened inference, which is supplemented with a MoE module to address the efficiency challenge. 
    \item We implement our solution on a well-known feature-based recommender Neural Factorization Machine (NFM)~\cite{NFM} and conduct extensive experiments on two real-world datasets, verifying the effectiveness of our proposal.
\end{itemize}


\section{Problem Definition}
We use an uppercase character (\eg $X$), calligraphic font (\eg $\mathcal{X}$) and lowercase character (\eg $x$) to denote a random variable, the sample space of the variable, and a specific value of the variable\zy{, respectively}. Let $\mathcal{D}$ denote historical interactions (\eg finished playing or click) where the user-item pairs are given binary labels. The target of recommendation is learning a model from $\mathcal{D}$ for predicting to what extent an item will match the user preference. Both user and item could be described by rich side information, \eg user demographics and item attributes (category, tags, \etc). 
Different from conventional settings~\cite{NFM}, we discriminate the \textit{confounding feature} ($A$) of item from the remaining \textit{content feature} ($X$). 

Conceptually, confounding feature is an item feature that has direct effect on the interaction $Y$, regardless of user preference. 
%
\noindent We assume that the confounding feature is a discrete variable of $K$ ($= |\mathcal{A}|$) values.  Because most item features are discrete in recommender system, and it is common to discretize continuous features for better modeling and interpretability.
Remarkably, there exist techniques~\cite{CD-Reinforcement,CD-zhangkun} for identifying such confounding feature. Thus,
this work focuses on 
the problem that with confounding feature as  known (\ie assuming the confounding feature has been identified), how to eliminate its impact in recommendation. As the first step, this work considers single confounding feature, and leaves the extension to multiple confounding features for future work.  
\section{Method}
We first present the causal view for analyzing the impact of the confounding feature on the recommendation process (Section~\ref{sec:causal-view}). Then we introduce our Deconfounding Causal Recommendation (DCR) framework  
to approach the user-item matching through backdoor adjustment (Section~\ref{sec:DCR-framework}).
Thirdly, we present a MoE model architecture to achieve efficient backdoor adjustment (Section~\ref{sec:MoE-architecture}). Lastly, we discuss the generality of DCR and show it retains the ability of addressing the confounding feature issue when causal relationship changes (Section~\ref{sec:dcr-generality}).

\subsection{Causal View of Confounding Feature}\label{sec:causal-view}

Conceptually, causal graph~\cite{pearl2009causality} is a directed acyclic graph, in which a node represents a variable and a directed edge denotes the causal relation between two connected nodes. Functionally speaking, causal graph is an abstract of the data generation process, and widely used to guide the model design~\cite{PDA,clickbait}. 
Figure~\ref{fig:our_graph_z} shows the causal graph for interaction generation when the confounding feature exists. 
Next, we explain the semantics of the causal graph. 

\begin{itemize}
   \item Node $U$ denotes the user, specifically, user features, including ID.
   \item Node $A$ denotes the given confounding feature.
   \item Node $X$ denotes \zy{other} content features of item, including ID, that are associated with user preference.
   \item Node $Z$ represents some hidden factors affecting both confounding feature $A$ and content feature $X$. Such factors are 
   caused by the 
   production of items, which 
   might be unobservable, 
   \eg the intention of the video creator.
   \item Node $M$ represents the user-item matching, reflecting to what extent the content features match user preference.
   \item Node $Y$ denotes the interaction label, indicating whether the interaction behavior (\eg \zy{finished playing} and click) is happened. 
  \item Edges $\{X, U\} \longrightarrow M$ denote that the user-item matching is determined by the user features $U$ and item content features $X$.
   \item Edges $\{A, M\}\longrightarrow Y$ represent that the interaction is determined both by the level of user-item matching $M$ and the confounding feature $A$. 
   The edge $A\longrightarrow Y$ corresponds to the phenomenon that the confounding feature \zy{affects} the probability of an interaction, but not reflects true user preference. 
  \item The edges $A \longleftarrow Z \longrightarrow X$ denote that hidden factor $Z$ affects both confounding feature $A$ and content features $X$. For instance, to create a wedding video, the creator ($Z$) will make relatively longer video ($A$) and choose touching background musics ($X$). As such, $A$ and $X$ will exhibit correlation in the observed data due to the hidden common cause $Z$.
\end{itemize}
Given the target of learning a ranking function from historical interactions $\mathcal{D}$, existing methods usually estimate the correlation $P(Y|U, X, A)$ or $P(Y|U, X)$. 
However, both choices are problematic: 
\begin{itemize}
    \item \textbf{Modeling} $P(Y|U,X,A)$. The models based on $P(Y|U,X,A)$ will account for the direct effect of $A \longrightarrow Y$. Consequently, the predictions are biased towards \zy{some special} $A$, \eg assigning higher scores for short videos.
    \item \textbf{Modeling} $P(Y|U,X)$. From the causal graph, we recognize a backdoor path between $X$ and $Y$, \ie $X \longleftarrow Z \longrightarrow A \longrightarrow Y$, wherein $A$ is a confounder between $X$ and $Y$. Therefore, ignoring $A$ in the input will make the model learn the spurious correlation between $X$ and $Y$, which also leads to biased recommendation. 
\end{itemize}

In this light, the key to eliminate the impact of $A$ lies in cutting off both the direct path $A \longrightarrow Y$ and backdoor path $X \longleftarrow Z \longrightarrow A \longrightarrow Y$ in the model prediction. Accordingly, \textit{making recommendation with the causal effect of $X$ on $Y$} can achieve the target.

\subsection{Deconfounding Causal Recommendation} \label{sec:DCR-framework}
We now consider how to obtain a recommender model based on the causal effect of $X$ on $Y$.

\subsubsection{Causal Intervention}
By definition, the causal effect of $X$ on $Y$ is the changes of $Y$ when forcibly changing the value of $X$ from a reference value to a target value. Therefore, the key to estimate causal effect lies in obtaining the outcome $Y$ after the intervention on $X$. In practice, a \textit{de facto} standard to obtain the outcome of causal intervention is conducting randomized controlled trial~\cite{RCT}. However, such experiments are very expensive in recommendation and not practical for the confounding feature issue, since the items are typically created by a third party. Therefore, we have to estimate the intervention outcome from the observational data.

\vspace{+5pt}
\noindent 
\textbf{Intervention with \textit{do-calculus}}. 
Fortunately, the \textit{do-calculus} in causal science~\cite{pearl2009causality} provides an alternative solution 
to estimate $P(Y | U, do(X))$.
\zy{Considering that there is a backdoor path between $X$ and $Y$, we take \textit{the backdoor adjustment}~\cite{pearl2009causality} to identify the target causal effect. Formally, we have,}
\begin{equation} 
    \label{eq:causal-effect}
    P(Y|U,do(X)) = \sum_{a\in \mathcal{A}} P(Y|U,X,A=a)P(A=a),
\end{equation}
where $P(Y|U,X,A=a)$ and $P(A=a)$ are both identifiable conditional probability distributions. 
Conceptually, $P(Y|U,do(X))$ blocks the backdoor path by conditioning on the confounder $A$. 
\zy{From the view of controlled experiment, it means that we select a group of candidate user-item pairs such that $A$ has the fixed distribution $P(A)$ across different groups, and set their $X$ as the target value; and then observe the outcome distribution $P(Y|U,X,A=a)$ over the candidates.}

Given the target value $x$ and reference value $x^*$, the difference between $P(Y|U,do(X=x))$ and $P(Y|U,do(X=x^{*}))$ is equal to the difference between the controlled experiments over two randomized groups since the marginal distribution $P(A)$ is invariant.
In this way, the obtained causal effect on $Y$ of changing the value of $X$ from the reference value $x^*$ to the target value $x$, is freed from the influence of the confounding feature $A$.
\zy{Remarkably}, we can directly regard $P(Y|U,do(X=x))$ as the causal effect of $X$ on $Y$, \ie discarding the reference status, since the reference status is \zy{the} same for all items and thus has no influence on item ranking. 
\zy{Theoretically, we can also conduct the backdoor adjustment over $Z$ as $P(Y|U,do(X)) = \sum_{z\in \mathcal{Z}} P(Y|U,X,Z=z)P(Z=z)$. Nevertheless, this is impractical since $Z$ is a hidden confounder that cannot be observed.}

\subsubsection{Estimating $P(Y|U,do(X))$.}
To estimate $P(Y|U,do(X))$, \zy{as the procedure in PD~\cite{PDA},} we need to: 1) model the two probability distributions in Equation~(\ref{eq:causal-effect}), \ie $P(Y|U,X,A)$ and $P(A)$, through historical interaction data $\mathcal{D}$; and 2) infer the expectation of $P(Y|U,X,A)$ over $P(A)$.

\vspace{+5pt}
\noindent \textbf{Estimating} $P(A)$. Recall that the confounding feature $A$ is discrete with $K$ possible values. We have $K \ll |\mathcal{D}|$, where $|\mathcal{D}|$ is the size of dataset $\mathcal{D}$. Therefore, we can directly approximate $P(A=a)$ with the ratio of samples with $A=a$, \ie 
\begin{equation}
    \label{eq:probability_A}
    P(A=a) = \frac{|\{ (U,X,A,Y) | A = a\}|}{|\mathcal{D}|}.
\end{equation}

\noindent \textbf{Estimating} $P(Y|U, X, A)$. 
Apparently, it is infeasible to directly observe all probabilities from $\mathcal{D}$ due to the data sparsity in recommendation. To resolve this issue, we resort to a machine learning model to learn the distribution. In line with existing recommendation work~\cite{jiawei2018modeling,sato2020unbiased}, we assume that $P(Y|U, X, A)$ follows a Bernoulli distribution.
Given a specific condition $U = u$, $X = x$, and $A = a$, we learn a mapping function $f(u, x, a)$ to calculate the interaction probability\footnote{Note that the output can be seen as the parameter of the Bernoulli distribution.} (\ie $Y=1$) by minimizing its negative log-likelihood over $\mathcal{D}$. Formally, 
\begin{align}\label{eq:training_obj}
    \min \sum_{(u,x,a,y)\in \mathcal{D}} &-y\log\left(
        f(u, x, a)
    \right)\\ \notag
    &-(1-y)\log\left(
        1-f(u, x, a)
    \right).
\end{align}
The $L_2$ regularization is used over the parameters of the mapping function to control overfitting but is not shown for briefness. We can implement the mapping function $f(u, x, a)$ by any feature-aware recommender model, like FM~\cite{RendleFM} and NFM~\cite{NFM}. 

\vspace{+5pt}
\noindent\textbf{Inference}. Once the model is trained, we can evaluate $P(Y|U,do(X))$ according to Equation~\eqref{eq:causal-effect} for recommendation scoring. Given a user-item pair, the recommendation score is calculated as: 
\begin{equation} \label{eq:result}
     {P} (y=1 | u, do(x)) = \sum_{a \in \mathcal{A}} P(a) \cdot f(u,x,a).
\end{equation}
To summarize, we first train a model according to Equation~(\ref{eq:training_obj}) to learn $P(Y|U,X,A)$. In the inference stage, we calculate ${P} (y=1|u, do(x))$ according to Equation~(\ref{eq:result}) to free the recommendation from the impact of the confounding feature $A$. We term this  general framework as \textit{Deconfounding Causal Recommendation} (DCR).

\begin{figure}
    \centering
    \includegraphics[width=0.66\textwidth]{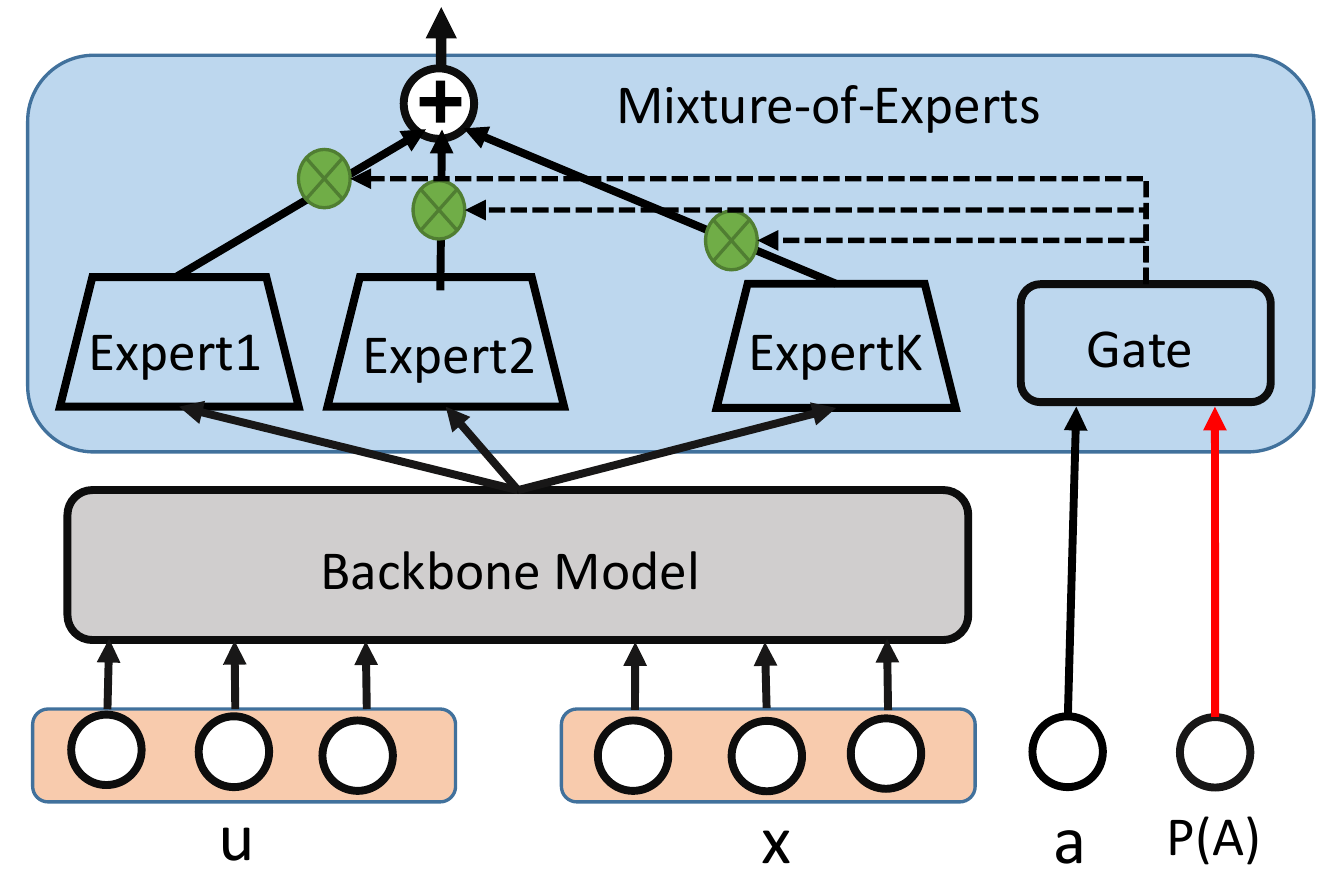}
    \caption{Illustration of the MoE model architecture for Deconfounding Causal Recommendation. 
    In the gate module, the inputs with black arrow and red arrow are used for model training and model inference, respectively.
    }
    \label{fig:model}
\end{figure}

\subsection{Mixture-of-Experts  Model Architecture} \label{sec:MoE-architecture}
As above, a direct way to evaluate ${P} (y=1 | u, do(x))$ is enumerating each value of $A$, conducting the model inference $f(u, x, a)$ for $K$ times (note that $K = |\mathcal{A}|$). 
Apparently, it will significantly increase the time cost in recommendation inference. 
As such, we need to consider how to accelerate the inference. 

\vspace{+5pt}
\noindent \textbf{NWGM approximation}. Note that ${P} (y=1 | u, do(x))$ is indeed the expectation of $ f(u,x,a)$ over the distribution $P(A)$. We can thus achieve the target with the NWGM approximation~\cite{xu2015show}, which has been widely used to approximate the expectation of functions~\cite{wenjiekdd,PDA,qi2020two}. Formally, 
\begin{align}\label{eq:nwgm} 
    {P} (y=1 | u, do(x)) \approx f\left(
        u, x, \sum_{a \in \mathcal{A}} a * P(a)
    \right).
\end{align}
Nevertheless, this approximation will sacrifice the estimation precision especially when the function $f(\cdot)$ is nonlinear.

Noticing that directly adjusting the inference strategy results in  the dilemma between efficiency and accuracy, we now consider to address the issue by adjusting the model architecture. We propose a MoE model framework with two considerations: 1) calculating the causal effect with one time of model inference; and 2) 
preserving sufficient modeling fidelity. The key lies in simultaneously calculating $f(u, x, a)$ for all values of $A$ in one model inference.
In this light, the MoE framework 
handles each value of $A$ with a specific expert\footnote{Each expert models a stratification~\cite{Principal_stratification} $P(Y|U,X,A)$ \wrt the value of $A$.}. We then let the experts share the same backbone for handling the matching between $U$ and $X$. This can largely save the computation cost since the $U$-$X$ matching, especially when considering high-order feature interactions~\cite{xDeepFM}, is the most computation-intensive part of $f(u, x, a)$. 

Concretely, the MoE framework (\cf Figure~\ref{fig:model}) includes: 
\begin{itemize}
    \item \textit{A Backbone Recommender Model}, which aims to learn a representation for the matching between item features $x$ and $u$:
\begin{equation} \label{eq:rec-backbone} 
         \bm{m} = f_{\Theta}(u,x),
\end{equation}
\zy{where $\bm{m}$ is a 
latent representation that denotes the matching signal, and $\Theta$ denotes the model parameters.}

    \item \textit{$K$ Experts}, where each expert corresponds to a confounding feature value $a$, mapping the matching representation $\bm{m}$ and $a$ to the interaction probability. We implement each expert as a multi-layer perceptron with two hidden layers. 
    Formally, 
    \begin{equation}  \label{eq:experts}
         f(u, x, a) = f_{\phi_a} (\bm{m} | a), 
    \end{equation}
    where $\phi_a$ denotes the parameters of the expert for $a$. 
    Across the experts, we use a gate with a one-hot input to select the expert for each training sample. In the inference stage, we feed $P(A)$ into the gate to calculate the weighted sum over the experts, which evaluates the causal effect $P(y=1 | u, do(x))$.
\end{itemize}

\zytois{We term the DCR implemented with the MoE model architecture as DCR-MoE. Algorithm~\ref{alg:DCR-MoE} shows the procedure of training (line 2-8) and inference (line 10-14) of DCR-MoE. In training, after the MoE has been initialized (line 2), we will iteratively update the MoE model in a mini-batch manner. Each iteration has three key steps: compute the matching $\bm{m}$ between item feature $x$ and $u$ with the same backbone recommender $f_{\Theta}$ for all samples (line 5); then compute the output of the expert corresponded to confounding value $a$, \ie $f_{\phi_{a}}(\bm{m}|a)$, for each sample sample $(u,x,a)$ (line 6); next update all model parameters by minimizing the loss in equation~\eqref{eq:training_obj} with   $f_{\phi_{a}}(\bm{m}|a)$ as the prediction for each sample (line 7). Note that although all experts are expected to be updated in each iteration, each training sample will only be fed into and used to update an expert that corresponds to the value of its confounding feature. During inference, we will first compute the matching $\bm{m}$ for each candidate user-item pair $(u,x)$ (line 11), then compute the outputs of all experts based on the $\bm{m}$ (line 12). Last, the prediction is generated by summing over all experts' outputs with $P(A)$ as weights (line 13). Different from the training, all experts will be utilized to generate the prediction for each candidate.}

\begin{algorithm}[t]
	\caption{DCR-MoE}
	\LinesNumbered
	\label{alg:DCR-MoE}
	\KwIn{Estimated $P(A)$ with equation~\eqref{eq:probability_A}, training dataset $\mathcal{D}$, testing dataset $\mathcal{D}_{testing}$, and the number of experts $K$}
	\KwOut{Predictions to user-item candidates in the testing dataset}
    // start training\;
    Initialize the the backbone recommender model $f_\Theta$ and the $K$ experts $\{f_{\phi_{a}}\}_{a\in\mathcal{A}}$ of the MoE\;
    \For{stop condition is not reached}{
    	Randomly sample a batch of data from $\mathcal{D}$ \;
    	Compute $\bm{m}=f_{\Theta}(u,x)$ for each training sample $(u,x,a)$ according to equation~\eqref{eq:rec-backbone}\;
    	For each sample $(u,x,a)$, compute the output of \textbf{the expert for} $a$, \ie $f_{\phi_{a}}(\bm{m}|a)$ in equation \eqref{eq:experts}\; 
    	Update the backbone recommender $f_\Theta$ and experts $\{f_{\phi_{a}}\}_{a\in\mathcal{A}}$ by minimizing the loss in equation~\eqref{eq:training_obj} with $f(u,x,a)=f_{\phi_{a}}(\bm{m}|a)$\; 
    }
    //start inference\;
    \For{ each candidate user-item pair $(u,x)$ in $\mathcal{D}_{testing}$}{
    Compute $\bm{m}=f_\Theta(u,x)$\;
    Compute the outputs of \textbf{all experts}, getting $\{f_{\phi_{a}}(\bm{m}|a)\}_{a\in\mathcal{A}}$\;
    Compute $\sum_{a\in\mathcal{A}} f_{\phi_{a}}(\bm{m}|a)P(a)$ as the prediction\;
    }
	return all predictions for testing dataset\;
\end{algorithm}

\subsection{Generality of DCR} \label{sec:dcr-generality}



\begin{figure} 
\centering
\subfigure[ \textbf{$A\longrightarrow X$}]{ \includegraphics[width=0.323\textwidth]{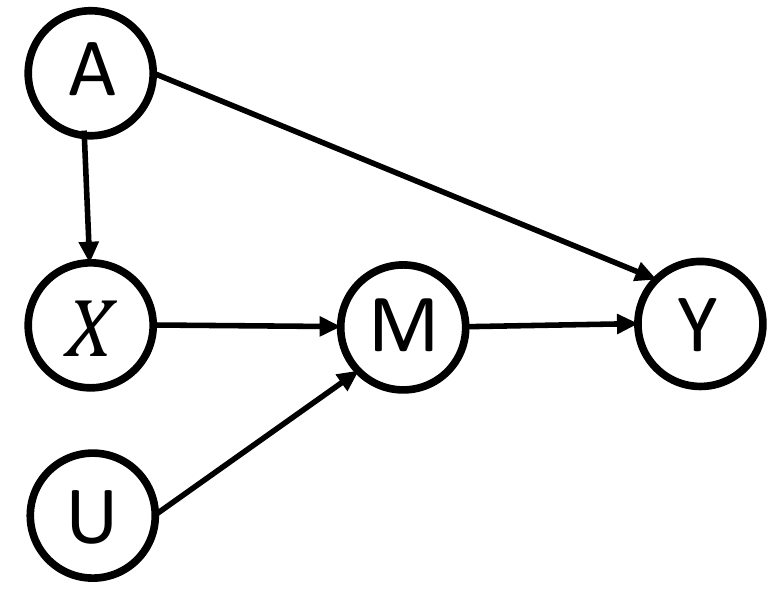}\label{subfig:case2}}
\subfigure[ \textbf{$X\longrightarrow A$}]{ \includegraphics[width=0.323\textwidth]{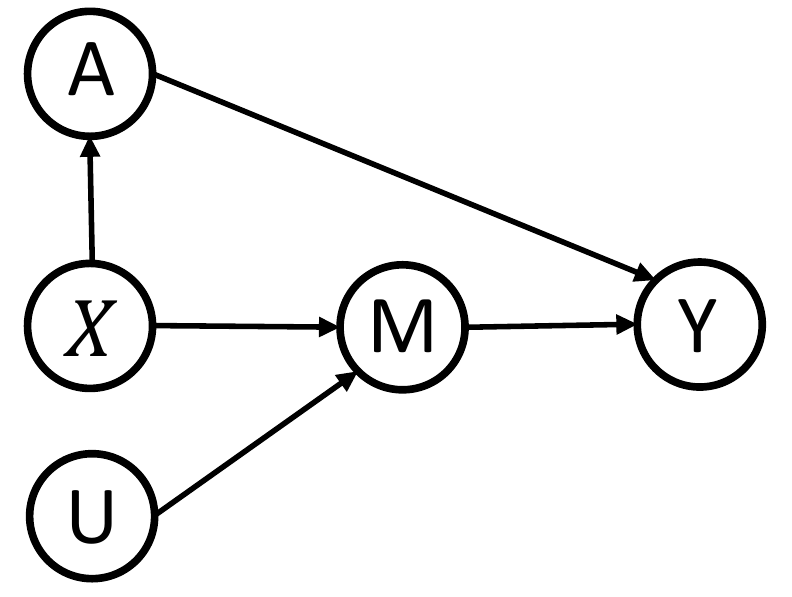}\label{subfig:case3}}
\subfigure[\textbf{No relation}]{\includegraphics[width=0.323\textwidth]{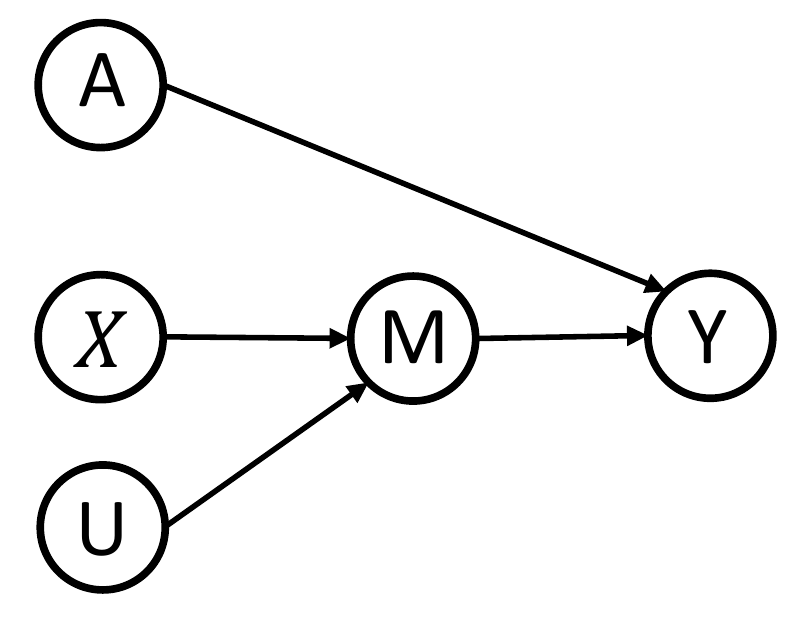}\label{subfig:case1}}
\caption{Three possibilities of direct causal relations between $A$ and $X$. Our DCR solution can also handle the three cases.}
\label{fig:more-graph}
\end{figure}

It is worth mentioning that DCR presents a general solution for addressing the confounding feature issue. It works not only for the case that a confounder exists between $A$ and $X$ (as shown in Figure~\ref{fig:our_graph_z}), but also for other cases that the causal relation between $A$ and $X$ is different. Considering the direct causal relation (i.e., no other variable like mediator or confounder) between $A$ and $X$, there are three possible cases:
\begin{itemize}
    \item $A \longrightarrow X$. As shown in Figure~\ref{subfig:case2}, $A$ has direct causal effect on $X$. In this case,
    $A$ is still a confounder between $X$ and $Y$. Our intervention with \textit{do-calculus} (abbreviated as \textit{do-intervention}) in Equation~\eqref{eq:causal-effect} still gives the causal effect of $X$ on $Y$. 
    
    \item $X \longrightarrow A$. 
    As shown in Figure~\ref{subfig:case3}, there is a directed edge from $X$ to $A$. 
    In this case, $X \longrightarrow A  \longrightarrow Y$ is also a causal path, \ie $A$ becomes one of the mediators for the causal effect of $X$ on $Y$. 
    It is thus harder to block the undesired direct effect of $A$ on $Y$, which cannot be achieved by estimating the total causal effect of $X$ on $Y$. This is because $P(Y|U,do(X))$ contains the effect of $A$ on $Y$.  To achieve the goal, we should dive into the path-specific causal effect of $X$ on $Y$ through the path $X\longrightarrow M \longrightarrow Y$. According to~\cite{pearl2009causality}, we define such causal effect as:  
    \begin{equation}
    \begin{split}
        &~~~P(Y|U,X=x,A_{x^*}) -  P(Y|U,X=x^{*},A_{x^*}) \\ 
        = & \sum_{a \in \mathcal{A}} \left(P(Y|U,x,a) - P(Y|U,x^{*},a)
        \right)P(a|x^{*}),
    \end{split}\label{eq:path_specific_ce}
    \end{equation}
    where $x$ and $x^{*}$ represent the target and reference values of $X$ respectively, and $A_{x^*}$ denotes whatever possible values of $A$ if let $X=x^{*}$. 
    All possible values of $x$ can take the same reference value $x^*$ to estimate their causal effects. 
    The second term in Equation~\eqref{eq:path_specific_ce} will thus not affect the ranking of items when using this path-specific causal effect as a recommendation policy. Therefore, our target is to estimate:
    \begin{equation}\small
        \sum_{a \in \mathcal{A}} P(Y|U,x,a) P(a|x^{*}).
    \end{equation}
    Remarkably, we can take any value of $X$ as the reference status. As long as there exists a $x^*$ satisfying $P(a|x^{*})=P(a)$, we have,
    \begin{equation}
        \sum_{a \in \mathcal{A}} P(Y|U,x,a) P(a|x^{*}) \propto \sum_{a \in \mathcal{A}} P(Y|U,x,a)P(a).
    \end{equation}
    It means our do-intervention can to some extent approximate the path-specific causal effect under this case.

    \item No relation. As shown in Figure~\ref{subfig:case1}, $A$ and $X$ are independent item features.
    In this case, performing intervention equals directly inferring the plain correlation. Formally,
    \begin{equation}
        P(Y|U,do(X)) = \sum_{a \in \mathcal{A}} P(Y|U,X,a)P(a) = P (Y|U,X).
    \end{equation}
    As we consider the average causal effect instead of the individual causal effect, our do-intervention still captures the causal effect of $X$ on $Y$. That is, the do-intervention still focuses on  the user-item matching, eliminating the direct effect of $A$ on $Y$. 
    
\end{itemize}

To summarize, the proposed do-intervention can eliminate the direct (and confounding) effects of the identified item feature $A$ and recognize the true user-item matching whatever the \zytois{(direct)} causal relations between $A$ and $X$\footnote{\zytois{Note that although we only discuss the cases regarding the direct causal relations between $A$ and $X$. Indeed, the discussed three cases can represent many other complicated cases: 1) $A\longrightarrow X$ can represent the case that there are paths from $A$ to $X$; 2) $X\longrightarrow A$ can represent the case that there are paths from $X$ to $A$; 3) no relation can also represent the case that the path between $A$ and $X$ contains colliders, since the colliders indeed block this path, making $A$ and $X$ independent. And the corresponding discussions can be directly applied to these complicated cases.}}.
It means that DCR (including MoE) is a general framework for tackling the confounding feature issue not restricted to the causal graph in Figure~\ref{fig:our_graph_z}.
Moreover, these analyses 
indicate the potential of taking DCR as a uniform framework to deal with other issues in recommendation that can be abstracted as the causal graphs in Figure~\ref{fig:our_graph_z} and \ref{fig:more-graph}. For example, the popularity bias issue~\cite{PDA} and position bias issue~\cite{ips-position} can be described by Figure~\ref{subfig:case2} and Figure~\ref{subfig:case3}, respectively.

\section{Related Work}

\zy{
In this section, we discuss the work on recommendation fairness and bias, which are both relevant with our work in terms of eliminating the impact of some predefined factors. We also review  
causal recommendation methods from the technical perspective.}

\subsection{Fairness in Recommendation}
\zy{In recommendation, fairness~\cite{fairness-survey} is usually defined on some sensitive features such as user gender~\cite{www2021career} and user race~\cite{fairtensor}.} 
Towards the fairness goal, most efforts~\cite{www2021career,li2021user,sigir2021-cuaslfair,pairwise-fair,fairgo} try to remove the information of sensitive features from making recommendations. A line of research achieves this target through post-processing~\cite{leonhardt2018user,li2021user}, which forcibly re-ranks the items with heuristically defined fairness policies. 
\cite{beyond,sigir2021-cuaslfair,fairgo,chuhan-fair,fairtensor} try to remove the sensitive information by adding the fairness-aware loss into the training objectives, \eg \zytois{ \cite{fairtensor} constraints the representations of sensitive and nonsensitive features to be orthogonal; \cite{beyond} adds the loss defined based on fairness metrics such as the non-parity.} Among them, \cite{fairgo,chuhan-fair,sigir2021-cuaslfair,li2022fairgan} adopt adversarial training \zytois{to optimize the recommendation loss and fairness loss.}

In summary, these fairness-oriented methods try to remove all information related to the sensitive feature regardless its usefulness for estimating user preference, which may reduce the modeling fidelity. \zy{In contrast, we aim to make the recommendation free from the impact of the confounding feature, including its direct and confounding effects, without sacrifice of the information of other item features ($X$ in Figure~\ref{fig:our_graph_z}).}
What's more, the above methods are at the level of correlation instead of causality. 
\subsection{Bias in Recommendation}
Recommender systems also face various bias issues generated by different reasons~\cite{chen-survey}.
The position bias~\cite{ips-position} and clickbait~\cite{clickbait} issues are most related to the confounding feature issue. Position bias happens as users tend to click items in the front positions of recommendation lists. Differently, the considered confounding feature is one type of item features instead of context features, and the underlying causal mechanisms are also different.
The clickbait issue~\cite{clickbait} considers the bias generated by the effect of the item's cover or title on click. Differently, we do not restrict the type of interaction. 
Moreover, its causal mechanism is different from us, \eg it does not consider confounders.
To handle bias issues, the most widely considered method is IPW-based methods\footnote{Please note that IPW is also one type of causal methods. \zytoisnew{The doubly robust (DR) based method~\cite{xiao2022towards} is also based on IPW but adds an imputation model to improve model robustness.}}~\cite{ICML2016-ips,saito2020MNAR,ips-position,huang2022different,xiao2022towards}, \zytois{which adjust the training distribution by reweighting training samples with propensity scores}. However, the propensities are difficult to set properly, e.g., causing the high variance issues etc \cite{PDA,chen-survey}. \zytois{\cite{wan2022cross,zhou2021contrastive} respectively take a cross-pairwise loss and  a constructive loss to achieve unbiased recommendation. They indeed implicitly reweight training samples by controlling the sampling of negative samples in the loss.} Utilizing unbiased data  (collected from random exposure)~\cite{distillation,autodebias,Selection-uniform} \zytois{to guide the model learning is another type of methods}, however such unbiased data is at the expense of user experience and risky to collect. {Post-hoc re-ranking or model regularization methods are also designed~\cite{abdollahpouri2017controlling,zhu2021popularity}, but they are heuristically designed and lack theoretical foundations for effectiveness~\cite{PDA}.} \zytois{Different from these methods, we take causal adjustment --- the backdoor adjustment --- to remove the undesired effects of the confounding features.}


\subsection{Causal Methods for Recommendation}
Recently, some efforts try to introduce causal inference to recommendation. 
One line of research considers the confounding issue in recommendation~\cite{PDA,wenjiekdd,wang2020causal,sato2020unbiased,christakopoulou2020deconfounding,shang2019environment}. Among them, \cite{PDA,wenjiekdd,TKDE2022-backdoor} also adopt causal intervention, where PD~\cite{PDA} aims at solving the popularity bias problem, and DecRS~\cite{wenjiekdd} considers the bias amplification problem. Technically, PD takes a similar two-step procedure to us for causal effect estimation, but assumes the form of the effect of confounders and designs a partially linear model to accelerate the inference. Differently, the proposed DCR does not need this assumption and takes MoE to accelerate the inference, keeping the model's expressiveness. DecRS takes a different procedure for causal effect estimation with the NWGM approximation, which may sacrifice the estimation precision. DCR directly estimates causal effects with MoE, ensuring both model fidelity and efficiency. \zytois{\cite{TKDE2022-backdoor} adopts the backdoor adjustment to eliminate the bias introduced by heterogeneous information. But their method is specially designed based on heterogeneous information networks and the backdoor adjustment is used for useful information selection instead of directly generating recommendations.} Besides,  \cite{wang2020causal} learns substitutes for unobserved confounders by fitting exposure data, to solve the unobserved confounder problem. \zytois{\cite{latentSeqRec} learns substitutes for unobserved confounders based on user historical interactions in sequential recommendation, and achieves deconfounding via IPW.} \cite{sato2020unbiased,christakopoulou2020deconfounding} adjust data distribution to estimate causal effect with IPW methods. \cite{liu2021mitigating} solves the confounding problem with information bottleneck~\cite{liu2021mitigating}. These methods do not perform intervention with \textit{do-calculus}.

Another type is the counterfactual-based method. Work~\cite{MACR, clickbait} tries to utilize the counterfactual inference to estimate their target causal effects. \cite{clickbait} is to solve the clickbait issue, and \cite{MACR} is to eliminate the popularity bias.  Both of them need to estimate the total effect and then remove the undesired effect by comparing the factual and counterfactual worlds. \zytois{\cite{CRepresentation} focuses on out-of-distribution recommendation. It first learns the interaction generation process and uses counterfactual inference to mitigate the effect of out-of-date interactions. Different from these methods, we take causal intervention instead of counterfactual inference to estimate the target causal effect.} Other work~\cite{counterfactual-seq, CauseRec,mehrotra2020inferring,accent,Counterfactual-Reward} is less related to us. \cite{counterfactual-seq, CauseRec,mehrotra2020inferring,yang2021top} try to generate some counterfactual data and take these data to improve the model performance, \zytoisnew{\eg \cite{yang2021top} generates counterfactual data by simulating the recommendation process and takes the generated data to train a ranking model.} \cite{Counterfactual-Reward} proposes a counterfactual importance sampling for recommendation based on the bandit. \cite{accent} proposes a framework for finding counterfactual explanations for neural recommenders. \zytois{Though these works consider the recommendation from the causal perspective, they make use the counterfactual inference to generate explanations or interaction data. We take intervention with \textit{do-calculus} to estimate causal effects and generate recommendations based on the estimated results.}

\section{Experiments}
In this section, we conduct experiments to answer the following three questions:

\noindent\textbf{RQ1}: How is the performance of the proposed framework DCR implemented with MoE (denoted as DCR-MoE) compared with existing methods?

\noindent\textbf{RQ2}: \zy{How do the design choices affect the effectiveness and efficiency of DCR-MoE?}

\noindent\textbf{RQ3}: Has the proposed method effectively eliminated the impact of the confounding feature? 

\subsection{Experimental Setting}


\subsubsection{Datasets} \label{sec:dataset}
To evaluate the effectiveness of addressing the confounding feature issue, we require datasets with: 
1) biased training data affected directly by the confounding feature; and 
2) unbiased testing data \wrt the confounding feature, which can reflect true user interests. 
In this light, we select two video recommendation datasets, treating the finished playing and a post-playing feedback as the training label and testing label of each sample (user-item pair), respectively. \zytois{Remarkably, taking a label that is different from the training label as the testing label is a reasonable setting to evaluate debiasing results~\cite{clickbait}.}
As aforementioned, video length is a confounding feature in video recommendation.
Note that each sample has both training and testing labels, which are both binary.
For each dataset, we split all samples into training/validation/testing sets with a ratio of $6:2:2$. To avoid data leakage, we make sure each sample regardless of label types only occurs in either training or testing/validation sets.

1) \textbf{Kwai}: It is a short video recommendation dataset released in the Kuaishou User Interest Modeling Challenge\footnote{https://www.kuaishou.com/activity/uimc}.
\zy{It is a relatively large dataset in research, having $19,429,358$ interacted samples between $18,019$ users and $422,144$ videos. It has two types of interaction labels: finished playing and liking. The finished playing is taken as the training label, and the other is taken as the testing label.}
\zy{ We discretize the confounding feature, \ie the video length, such that it has 6 possible values. We take other items' discrete features that can be one-hot encoded as the other item feature $X$, indeed, only id information can be utilized.}

2) \textbf{Wechat}: \zy{The dataset is released in the WeChat Big Data Challenge\footnote{https://algo.weixin.qq.com/}, which records user behaviors on short videos. It has $7,195,486$ interacted samples between $20,000$ users and $96,539$ videos. It contains many types of interaction labels, such as finished playing, liking, and read-comment.
We take the finished playing as the training label, and the read-comment as the testing label.}
We also discretize the confounding feature (video length) such that it has 6 possible values. We take some other inherent and discrete features such as \textit{bgm\_song\_id} as the other item feature $X$. 

\zytois{To show that the selected testing label is more free from the impact of the confounding feature compared with the training label, we compute the Pearson correlation coefficients between the confounding feature and the two labels, respectively. The Pearson correlation coefficient is a measure of linear correlation between two variables\footnote{The measure can only reflect a linear correlation of variables, and ignores many other types of relationships or correlations. Anyway, it is one of the most widely used measurements.}. The correlation coefficient ranges from $-1$ to $1$, and a higher absolute value implies a stronger linear correlation of the variables. We denote the correlation coefficient between the confounding feature and the training label as $\rho_1$, and the correlation coefficient between the confounding feature and the testing label as $\rho_2$. The computed correlation coefficients and other statistics of datasets are summarized in Table~\ref{tab:data-info}. We can find that for both the two datasets, the absolute value of $\rho_2$ (\ie $|\rho_2|$) is far small than that of $\rho_1$ (\ie $|\rho_1|$), and $\rho_2$ is very close to zero. These results show that the testing label has weaker linear correlations to the confounding feature, implying the testing label is more free from the impact of the confounding feature to some degree. Therefore, we think taking the selected labels as testing labels is appropriate.}

\begin{table}[]
\caption{Dataset Information. $A$ denotes the confounding feature. $\rho_{1} (or \rho_{2})$ denotes the Pearson correlation coefficient between the confounding feature $A$ and the training (or testing) label.}
\label{tab:data-info}
\begin{tabular}{@{}cccccccc@{}}
\toprule
Dataset & \#users   & \#items    &\#samples & $A$ &$\rho_{1}$ & $\rho_{2}$ & $|\rho_{1}/\rho_{2}|$\\ \midrule
Kwai    & 18,019 & 422,144 & 19,429,358  &video length& -0.085     & 0.014  & 6.07  \\
Wechat  & 20,000 & 96,539  & 7,195,486   &video length&  -0.131     & -0.015 & 8.73 \\ \bottomrule
\end{tabular}
\end{table}



\subsubsection{Compared Methods}
We implement the DCR framework based on the MoE architecture, \ie DCR-MoE. Specially, we implement the backbone recommender model of MoE as the bottom layer of Neural Factorization Machines (NFM)~\cite{NFM}, including embedding and the bi-interaction (EB) layers, and implement each expert of MoE as the deep part of NFM, including hidden and prediction layers. Then, we compare it with the following correlation-based, IPW-based, fairness-oriented, and counterfactual inference based methods:


\noindent\textbf{-NFM-WA}~\cite{NFM}. 
This method trains NFM with the confounding feature as the input, \ie estimating the correlation $P(Y|U,X,A)$ as the user-item matching.

\noindent\textbf{-NFM-WOA}. This method trains NFM without the confounding feature as the input, \ie estimating the correlation $P(Y|U,X)$ as the user-item matching.


\noindent\textbf{-IPW}~\cite{saito2020MNAR}, refers to the conventional inverse propensity weighting method. \zy{It tries to capture true user preference from biased data by re-weighting training samples}. \zy{Following the previous work~\cite{saito2020MNAR}, 1) we assume the interaction probability is equal to $P(Y=1|A)P(R=1|U,I)$, where $R$ represents the relevance between user $U$ and item $I$; 2) the propensity weight is defined as the inverse of $(\frac{P(Y=1|A=a)}{\max_{a^{'}}{P(Y=1|A=a^{'})}} )^{0.5}$ for positive samples with $A=a$.} The weights for negative samples are defined in a similar way. We also implement IPW with NFM. 

\noindent\textbf{-FairGo}~\cite{fairgo}. \zy{This method is proposed for dealing with unfairness probelm in recommendation. To achieve fair recommendation, it attempts to remove the effects of sensitive features by adversarial training. Here, we take it to eliminate the impact of the confounding feature, \ie taking the confounding feature as the sensitive feature. We implement it based on NFM for a fair comparison. The hyper-parameter $\lambda$ to control removing the effects of the sensitive feature is tuned in the range of $\{1e\text{-}3,1e\text{-}2,1e\text{-}2,0.1,0.2,10,50\}$.}

\noindent\textbf{-CR}~\cite{clickbait}. This is a counterfactual method for clickbait issue~\cite{clickbait}. \zy{It trains the model with the exposure feature but removes the undesired effect of exposure features at inference \zytois{with counterfactual reasoning}. We replace the exposure feature with the confounding feature $A$ to eliminate the impact of $A$. We implement CR based on NFM, and take the \textit{SUM-tanh} strategy (showing the best results in~\cite{clickbait}) 
to fuse the effects of exposure features ($A$) and other features ($X$) on interactions.} The hyper-parameter $\alpha$ to control the influences of exposure feature is searched in the range of $\{0.1,0.25,0.5,0.75,1,2,3,4,5\}$.


\subsubsection{Hyper-parameters} For a fair comparison, \zytois{all methods are optimized with the binary cross-entropy (BCE) loss} and tuned on the validation set. We optimize all models with adagrad~\cite{adagrad} optimizer \zy{with default mini-batch size of 1024}. 
Following previous work~\cite{deepFM} to set a small embedding size for FM-based methods,
we fix the embedding size to $16$ for all compared methods. The deep part of NFM (including hidden layers and a prediction layer) and the experts of our MoE are implemented as MLPs having two hidden layers with sizes $256$ and $128$, respectively. We search the learning rate in the range of $\{0.01,0.001\}$. \zy{For all methods, we have two different $L_2$ regularization coefficients: one for embedding layers and the other for other model parameters, which are both searched in the range of $\{1e\text{-}1,1e\text{-}2,\dots,1e\text{-}6,0\}$.} The best hyper-parameters for reported results are found by the grid-search with a popular hyper-parameter tuning tool -- ray\footnote{\zytoisnew{The document can be found at https://docs.ray.io/en/master/tune/index.html. And we take its ASHAScheduler to perform the grid-search process.}}~\cite{ray}, and the patience for early stopping is set as $10$ epochs.

\subsubsection{Metrics} To measure the recommendation performance, we adopt three widely-used evaluation metrics: Recall, \zy{Mean Average Precision (MAP)}, which consider whether the relevant items are retrieved within the top-$N$ positions, and NDCG that measures the relative orders among positive and negative items in the top-$N$ list. We generate recommendation lists just in the observed data, \ie ranking all the items appearing in the validation or testing sets, \zy{similar to \cite{autodebias}}. Since the average number of interacted items \zytois{by a user} in Kwai is significantly bigger than that in Wechat, \zytois{we report the results of top-$10$ recommendation and top-$20$ recommendation for Kwai, and the results of top-$3$ recommendation and top-$5$ recommendation for Wechat.} 

\begin{table}[]
\caption{
\zytois{The overall top-N recommendation performance of different methods on Kwai and Wechat. \zytoisnew{Metric@N (Recall@10) denotes the corresponding top-N (top-10) recommendation performance on this metric (Recall)}. For each dataset, bold scores denote the best in each column, while the underlined scores denote the best baseline. “RI” refers to the relative improvement of DCR-MoE over the corresponding baseline, averaged on the three metrics. For all metrics, higher results are better.
}}
\label{tab:overall}
\resizebox{0.99\textwidth}{!}{
\begin{tabular}{ccccccccc}
\toprule
\multicolumn{9}{c}{Kwai}                                                                                                                       \\
Methods          & Recall@10       & MAP@10          & NDCG@10         & RI     & Recall@20       & MAP@20          & NDCG@20         & RI     \\ \hline
NFM-WA           & 0.0778          & 0.0247          & 0.0458          & 40.4\% & 0.1530          & 0.0310          & 0.0694          & 30.7\% \\
NFM-WOA          & 0.0759          & 0.0228          & 0.0439          & 47.6\% & 0.1494          & 0.0289          & 0.0672          & 36.4\% \\
IPW              & 0.0775          & 0.0250          & 0.0464          & 39.5\% & 0.1474          & 0.0308          & 0.0684          & 33.2\% \\
FairGo           & {\ul 0.0876}    & {\ul 0.0269}    & {\ul 0.0506}    & 26.9\% & {\ul 0.1651}    & {\ul 0.0332}    & {\ul 0.0749}    & 21.4\% \\
CR               & 0.0764          & 0.0244          & 0.0458          & 41.8\% & 0.1492          & 0.0305          & 0.0687          & 33.0\% \\
\textbf{DCR-MoE} & \textbf{0.1089} & \textbf{0.0353} & \textbf{0.0634} & -      & \textbf{0.1936} & \textbf{0.0423} & \textbf{0.0896} & -      \\
\bottomrule \toprule
\multicolumn{9}{c}{Wechat}                                                                                                                     \\
Methods          & Recall@3        & MAP@3           & NDCG@3          & RI     & Recall@5        & MAP@5           & NDCG@5          & RI     \\ \hline
NFM-WA           & 0.1275          & 0.0910          & 0.1198          & 6.5\%  & 0.1584          & 0.0906          & 0.1341          & 4.2\%  \\
NFM-WOA          & 0.1324          & {\ul 0.0950}    & {\ul 0.1243}    & 2.4\%  & {\ul 0.1609}    & {\ul 0.0930}    & {\ul 0.1370}    & 2.0\%  \\
IPW              & {\ul 0.1326}    & 0.0941          & 0.1240          & 2.8\%  & 0.1600          & 0.0926          & 0.1366          & 2.5\%  \\
FairGo           & 0.1291          & 0.0924          & 0.1213          & 5.0\%    & 0.1598          & 0.0813          & 0.1351          & 7.6\%  \\
CR               & 0.1282          & 0.0915          & 0.1205          & 5.9\%  & 0.1584          & 0.0905          & 0.1345          & 4.1\%  \\
\textbf{DCR-MoE} & \textbf{0.1355} & \textbf{0.0976} & \textbf{0.1271} & -      & \textbf{0.1646} & \textbf{0.0947} & \textbf{0.1396} & -      \\
\bottomrule
\end{tabular}}
\end{table}

\subsection{RQ1: Performance Comparison}

In this subsection, we study the recommendation performance of our DCR-MoE over all users as well as in different user subgroups. 

\subsubsection{Overall Performances.} The overall performance comparison is summarized in Table~\ref{tab:overall} \zy{regarding the top-$N$ recommendation}.
From the table, we have the following observations:

\begin{itemize}
    \item The proposed method DCR-MoE achieves the best performance on both Kwai and Wechat datasets. This verifies the effectiveness of our deconfounding causal recommendation framework. The improvements can be attributed to the intervention performed at inference, \zy{\ie estimating the casual effect $P(Y|U,do(X))$ as user-item matching,
    making the recommendations free from the impact of the confounding feature.}
    
    \item DCR-MoE consistently outperforms NFM-WOA and NFM-WA. \zy{NFM-WA models the correlation $P(Y|U,X,A)$, thus the direct effect of the confounding feature on interaction will be captured, misleading the model towards items with dominant values of the confounding feature (\cf{Figure~\ref{fig:avg-pre}})}. \zy{The unsatisfactory results of NFM-WOA, which models the correlation $P(Y|U,X)$ , demonstrate that simply removing the confounding feature during model training cannot solve the confounding feature issue. This is because the spurious correlation brought by the backdoor path will mislead the results. Summarily, when the confounding feature appears, it is better to model the user-item matching from the causal effect perspective instead of mere correlation.}

    
    \item IPW and CR are both causal methods. However, \zy{neither of them demonstrates the power to deal with the confounding feature issue due to two drawbacks. 1) IPW relies on accurate estimation of propensities, which is non-trivial due to high variance~\cite{PDA,chen-survey}. 
    Moreover, its way to define propensity might not be suitable for the confounding feature.
    } 
    2) \zy{To remove the undesired effect of the confounding feature}, CR needs to precisely disentangle the effects of the confounding feature and other features. \zy{However, disentangling is hard without supervision or inductive bias ~\cite{DICE,ICML2019best}. Meanwhile, \zytois{the fusion strategy of CR} may be 
    insufficient to represent the real fusion mechanism of the above two types of effects.}

    
    \begin{figure} 
\centering
\subfigure[\textbf{ Recall@10 on Kwai}]{\includegraphics[width=0.493\textwidth]{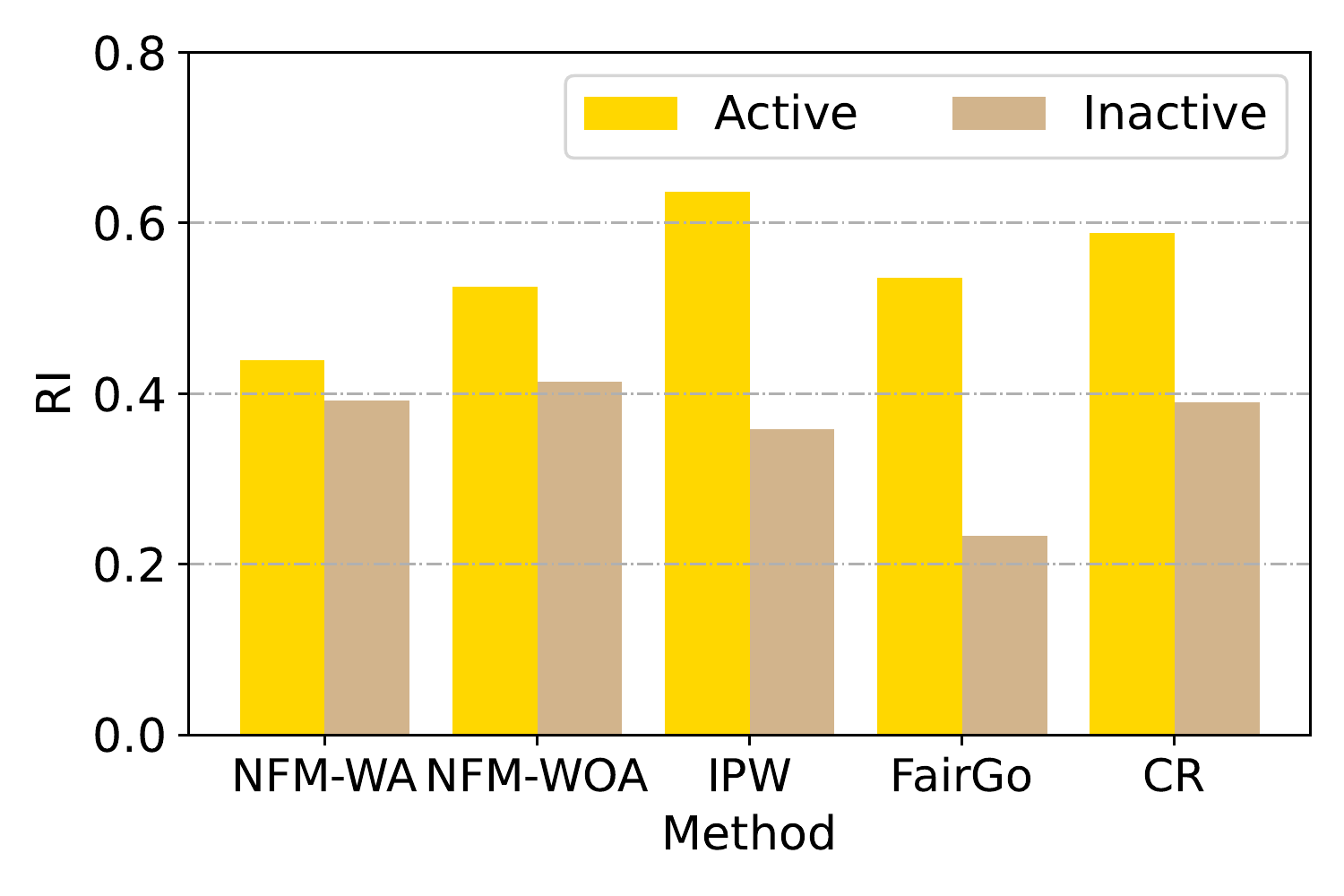}}
\subfigure[ \textbf{NDCG@10 on Kwai}]{ \includegraphics[width=0.493\textwidth]{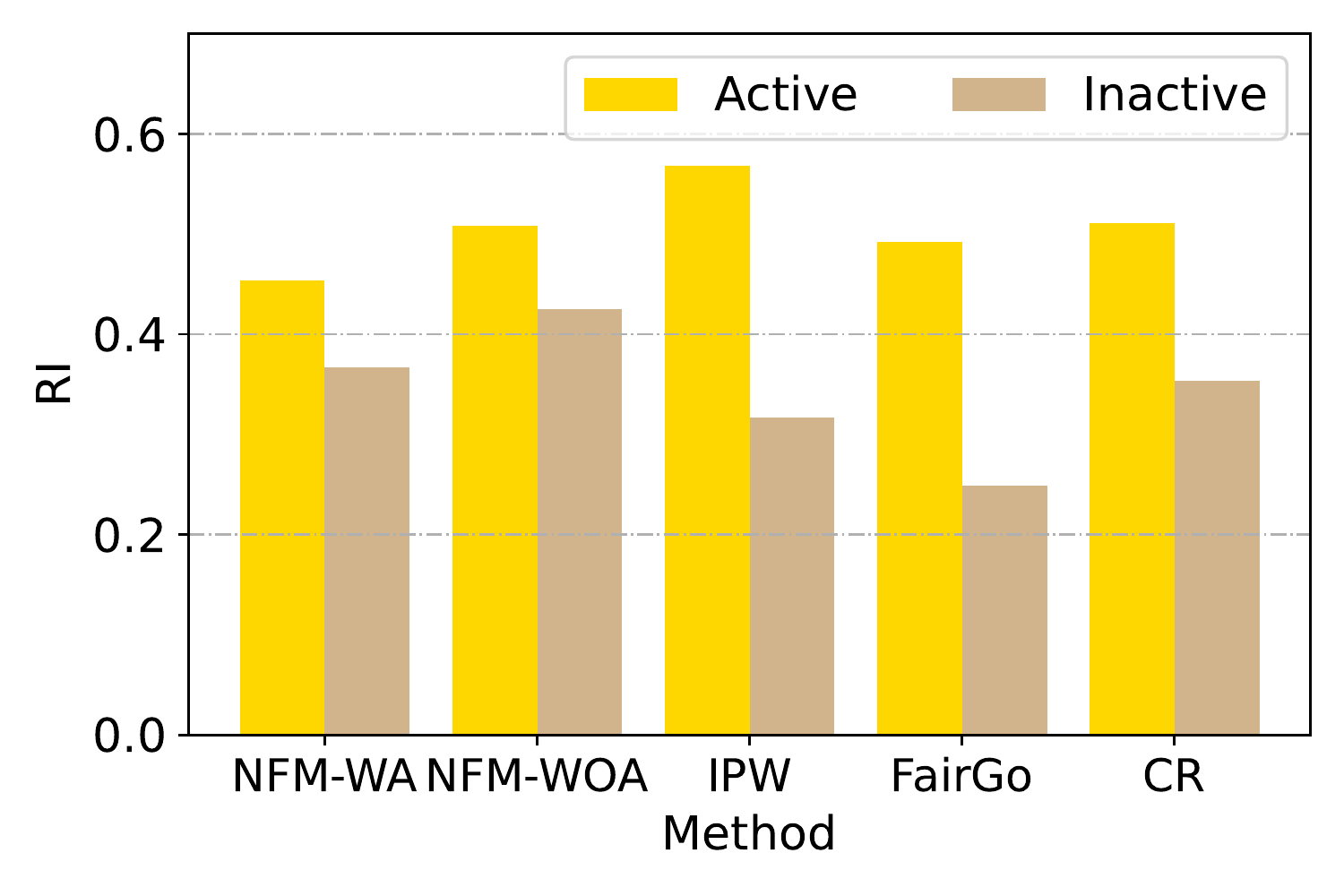}}
\subfigure[ \textbf{Recall@3 on Wechat}]{ \includegraphics[width=0.490\textwidth]{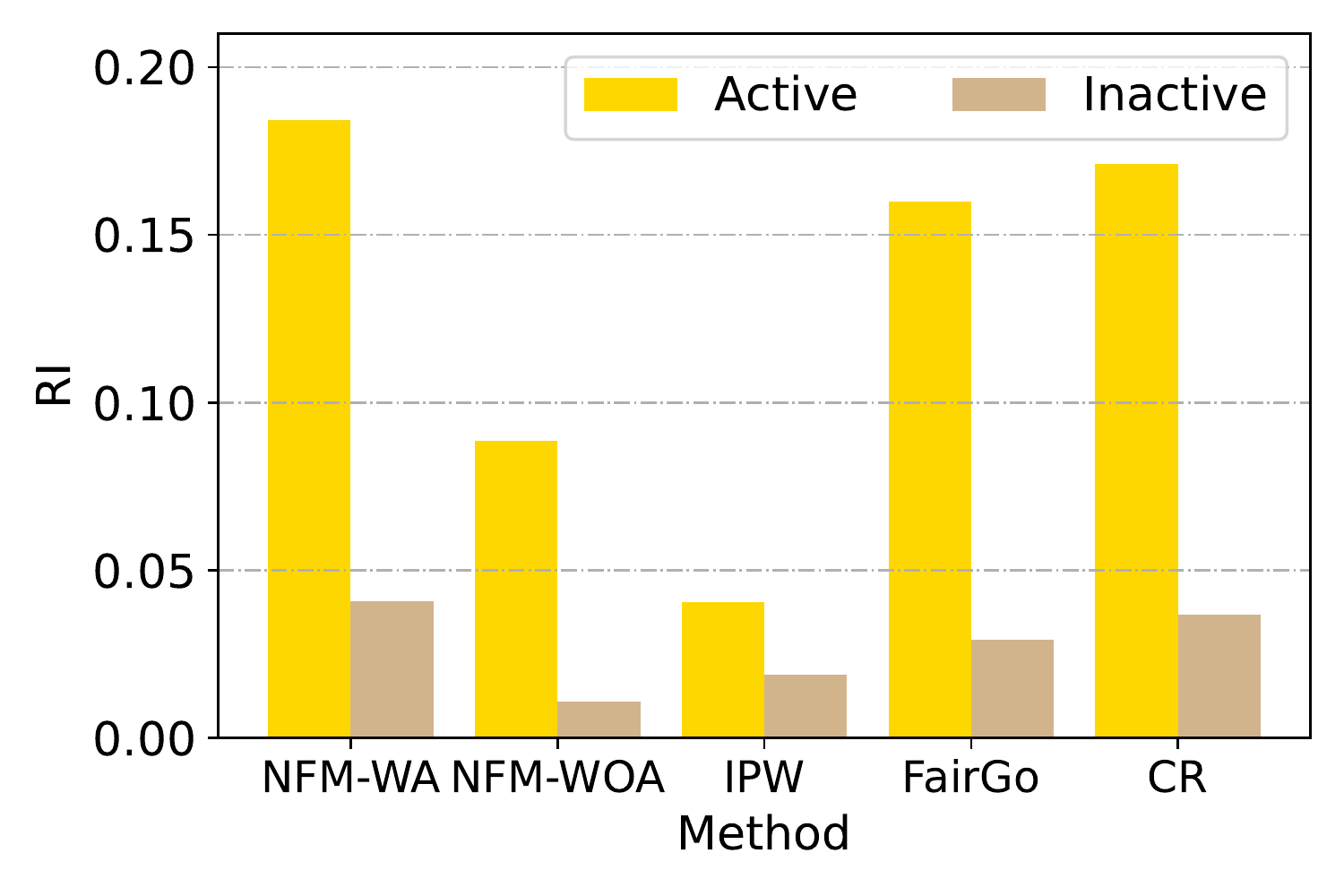}}
\subfigure[ \textbf{NDCG@3 on Wechat}]{ \includegraphics[width=0.490\textwidth]{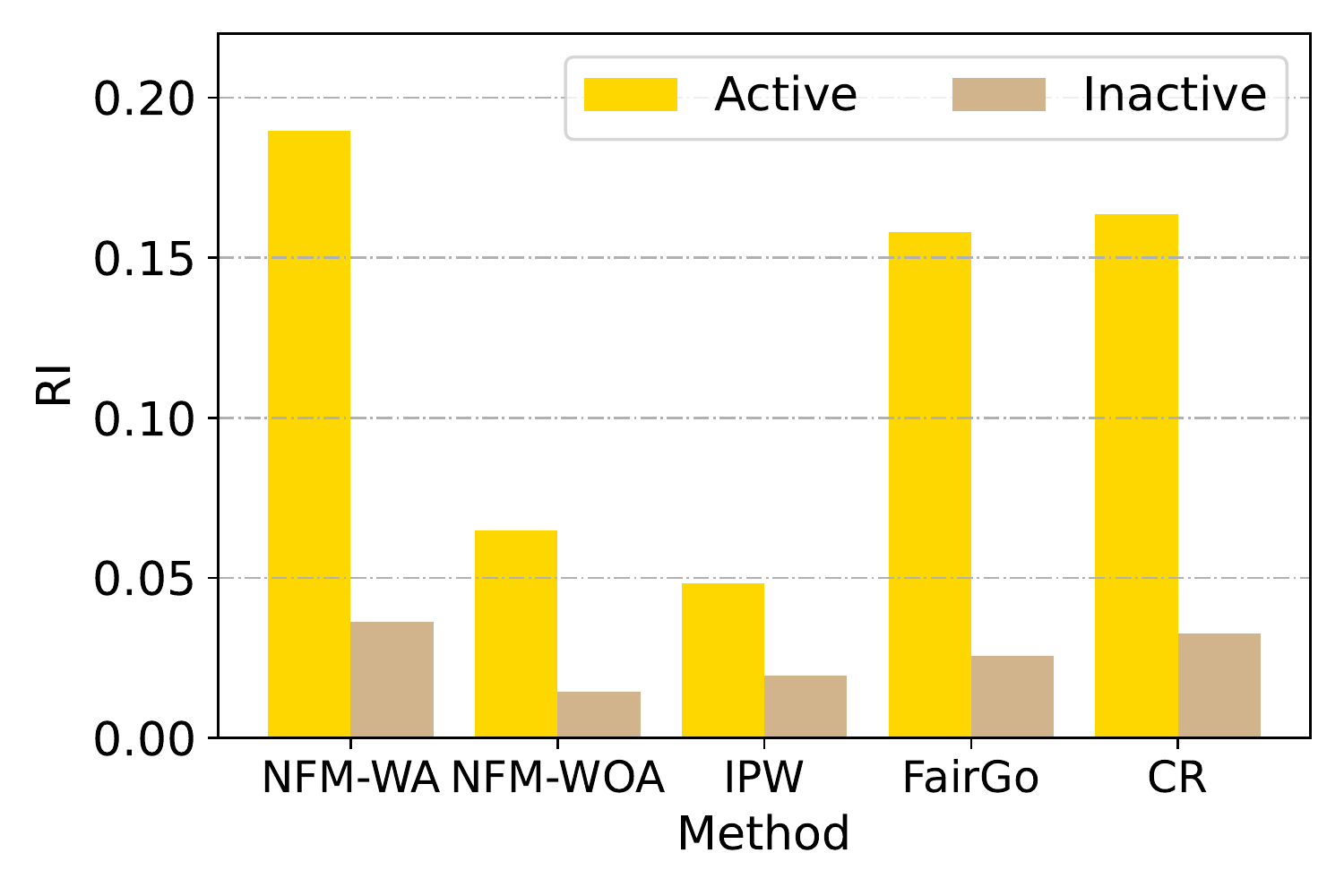}}
\caption{Relative Improvements(RI) of DCR-MoE over different baselines on the active and inactive user groups, respectively. \zytois{For Kwai, the results of Recall@10 and NFCG@10 are reported. For Wechat, the results of Recall@3 and NDCG@3 are reported. We omit the results of other metrics which show similar phenomena.}}
\label{fig:RI-groups}
\end{figure}
    
    \item The results of FairGo are also unsatisfactory where we postulate the reason to be information loss. The adversarial learning in FairGo tries to remove any information related to the confounding feature, \zy{which indeed removes partial information of $X$ due to the relations $X \longleftarrow Z \longrightarrow A$.} \zy{However, the removed information of $X$ may be useful for user-item matching.}  In contrast, DCR-MoE keeps all information of $X$ and merely eliminates the impact brought by the confounding feature.
    
    \item \zy{DCR-MoE achieves different degrees of improvements on Kwai and Wechat. We think the difference
    is due to different properties of the two datasets: 1) the interactions in Wechat are influenced by the social networks
    that are not given\footnote{\zytois{The dataset Wechat is a short video recommendation dataset collected from the China popular social media WeChat. Friends will influence each other to watch videos.}}; and 2)
    Wechat is more sparse regarding users.} 
\end{itemize}

\subsubsection{Improvements in Active and Inactive User Groups} 
\zytoisnew{We have partially attributed the performance improvement differences of DCR-MoE on Kwai and Wechat to the sparsity of the datasets on the user side.} We then investigate whether user activeness
affects the performance of our DCR-MoE. For this goal, we first split users into active and inactive groups according to the number of user's \zy{positive testing label} (\ie the positive post-playing feedback) and the number of user's training samples. 
\zy{The active group is the intersection of the following two user sets:} 1) users with the top-ranked number of \zy{training} samples (top $40\%$ for Kwai and top $20\%$  Wechat) and 2) users with the most positive testing labels (top $50\%$ for Kwai\footnote{In Kwai, each user in the top $50\%$ has great than 700 training samples.} and top $20\%$ for Wechat). \zy{Finally, the $20\%$ and $6\%$ of users are selected as  active users in Kwai and Wechat, respectively.}
Users that are not categorized as active users are all in the inactive group.
Then we evaluate different methods on the two groups respectively and compute the relative improvements (RI) of DCR-MoE over different methods. The results on Recall@10 and NDCG@10 for Kwai, on Recall@3 and NDCG@3 for Wechat, are shown in Figure~\ref{fig:RI-groups}. \zytoisnew{The results of other metrics are omitted since they show similar phenomena to the reported metrics.} 

According to the results in Figure~\ref{fig:RI-groups}, DCR-MoE can get improvements \zytois{in} both active and inactive groups. Moreover, DCR-MoE always achieves larger improvements \zytois{in} the active group. \zy{Particularly}, the improvements \zytois{in the active group of Wechat are at least twice as large as that in the inactive group.} These results reflect the influences of data sparsity on estimating causal effect: with more data, \zy{it is more possible that there are more diverse combinations of $X$ and $A$, thus $P(Y|U,X,A)$ can be estimated better on different values of $X$ and $A$ for user $U$; meanwhile, computing $P(Y|U,do(X))$ needs to enumerate all possible values of $A$; thus the estimation of $P(Y|U,do(X))$ regarding active users can be more accurate\footnote{Indeed, data sparsity is related to the \textit{overlap} assumption~\cite{blessings} for precise causal effect estimation.}.} \zytois{Second}, there are more inactive users than active users in both datasets, ecpecially for Wechat. The total relative improvements are dominated by the inactive users since the metrics are averaged over all users. This also explains why the overall performance gain on  Wechat dataset (\cf Table~\ref{tab:overall}) is relatively marginal. \zy{Besides}, we believe bringing more improvements for active users is also meaningful since the consumption of items is mostly generated by active users on real-world recommendation platforms.

\subsection{RQ2: Ablation Studies}
DCR-MoE has two main designs -- the intervention at inference and the MoE model architecture.  In this subsection, we conduct experiments to verify the function of these designs.

\begin{table}[]
\caption{The performance comparison between NFM-WA, MoE, and DCR-MoE. \zytois{Both NFM-WA and MoE estimate the correlation P(Y|U,X,A), but MoE takes the proposed MoE model architecture. MoE and DCR-MoE have the same model size.}}
\label{tab:do-calculus}
\resizebox{0.9\textwidth}{!}{
\begin{tabular}{c|ccc|ccc}
\hline
 & \multicolumn{3}{c|}{Kwai} & \multicolumn{3}{c}{Wechat} \\ \cline{2-7} 
\multirow{-2}{*}{\begin{tabular}[c]{@{}c@{}}Datasets\\ Methods\end{tabular}} & { { Recall@10}} & MAP@10 & NDCG@10 & { { Recall@3}} & MAP@3 & NDCG@3 \\ \hline
NFM-WA & 0.0778 & 0.0247 & 0.0458 & 0.1275 & 0.0910 & 0.1198 \\
MoE & 0.0757 & 0.0240 & 0.0448 & 0.1286 & 0.0924 & 0.1215 \\ 
DCR-MoE & 0.1089 & 0.0353 & 0.0634 & 0.1355 & 0.0976 & 0.1271 \\ \hline
\end{tabular}
}
\end{table}

\subsubsection{The Effectiveness of Intervention at Inference}\label{ab:intervention} Recall that the most important operation in our DCR framework is to perform intervention at the inference stage with \textit{do-calculus}. \zy{We have shown DCR-MoE can achieve better recommendation performances.} We question that the improvements of DCR-MoE may come from its more parameters instead of the intervention, because it has multiple experts. 
In this light, we further evaluate a variant of DCR-MoE, named MoE, which directly takes the expert-specific output of the MoE model, \ie still using a one-hot vector to select one expert at inference, 
as the recommendation score. Obviously, MoE models the correlation $P(Y|U,X,A)$ as the user-item matching.
Then we compare MoE with DCR-MoE, and \zy{NFM-WA that also models $P(Y|U, X, A)$ but with fewer model parameters.}
The comparison is shown in Table~\ref{tab:do-calculus}. \zytois{We see the performance of MoE is not as good as that of DCR-MoE which conducts intervention at inference.} \zy{And MoE has similar performances as NFM-WA \zytois{on the both Kwai and Wechat}, since they both estimate $P(Y|U,X,A)$ as user-item matching.} These results verify that our improvements are coming from the intervention \zytois{instead of the extra model parameters}. 

\begin{table}[]
\caption{Inference cost comparisons between DCR-NFM, DCR-NFM-A, and DCR-MoE (\zytois{the number of experts K=6}). Results shown in EB/MLP column give the number of times that different methods need to run EB layers/MLPs in NFM or MoE. The Time column refers to actual time cost (seconds) at inference.}
\label{tab:time-cost}
\resizebox{0.55\textwidth}{!}{
\begin{tabular}{c|ccc|ccc}
\hline
\multirow{2}{*}{\begin{tabular}[c]{@{}c@{}}Datasets\\ Methods\end{tabular}} & \multicolumn{3}{c|}{Kwai} & \multicolumn{3}{c}{Wechat} \\ \cline{2-7} 
 & EB & MLP & Time(s) & EB & MLP & Time(s) \\ \hline
DCR-NFM & 6 & 6 & 13.6 & 6 & 6 & 2.9 \\ 
DCR-NFM-A & 1 & 1 & 2.5 & 1 & 1 & 0.6 \\ 
DCR-MoE & 1 & 6 & 5.4 & 1 & 6 & 1.1 \\ \hline
\end{tabular}}
\end{table}

\begin{figure}
\centering
\subfigure[\textbf{ Kwai}]{\includegraphics[width=0.49\textwidth]{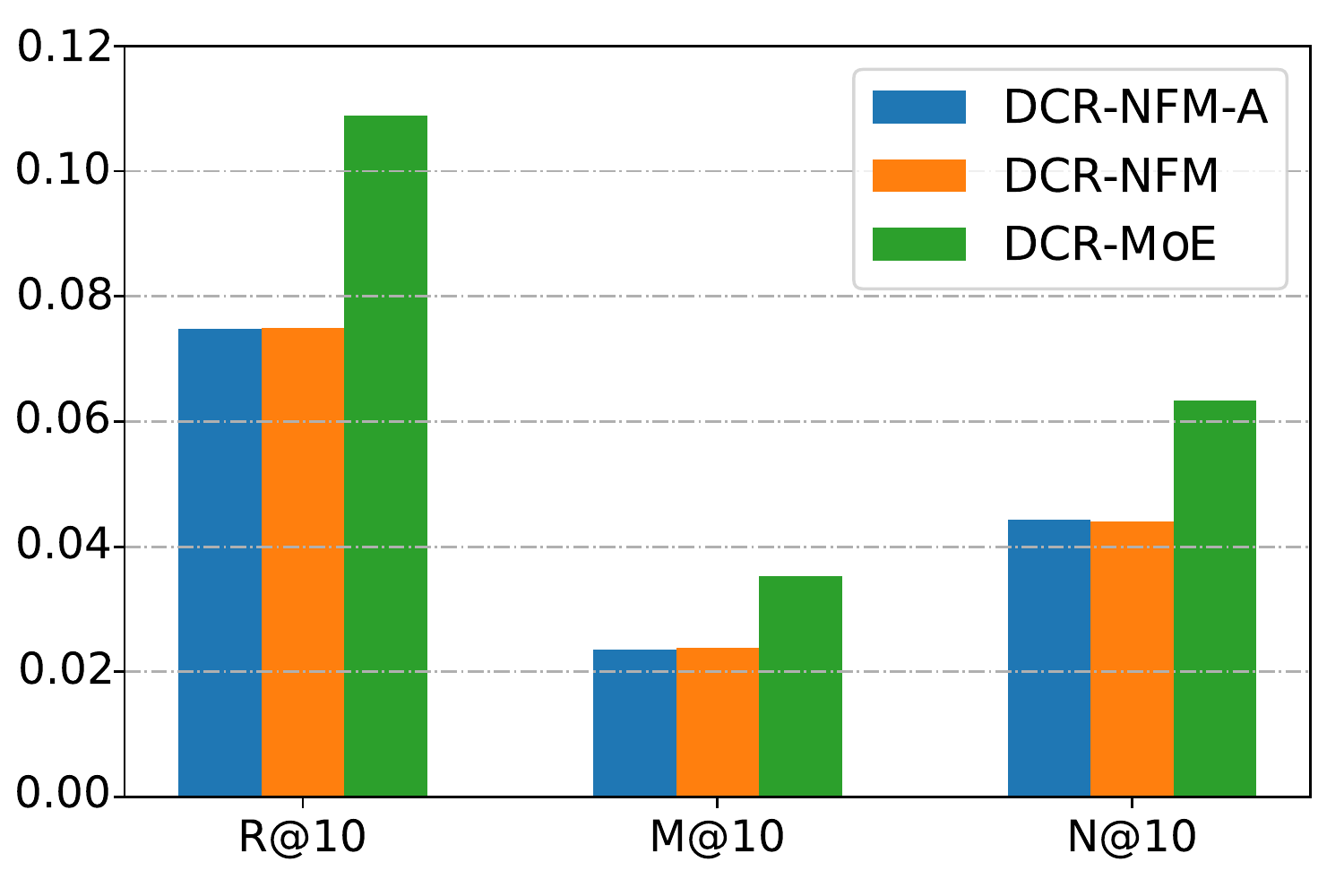}}
\subfigure[ \textbf{Wechat}]{ \includegraphics[width=0.49\textwidth]{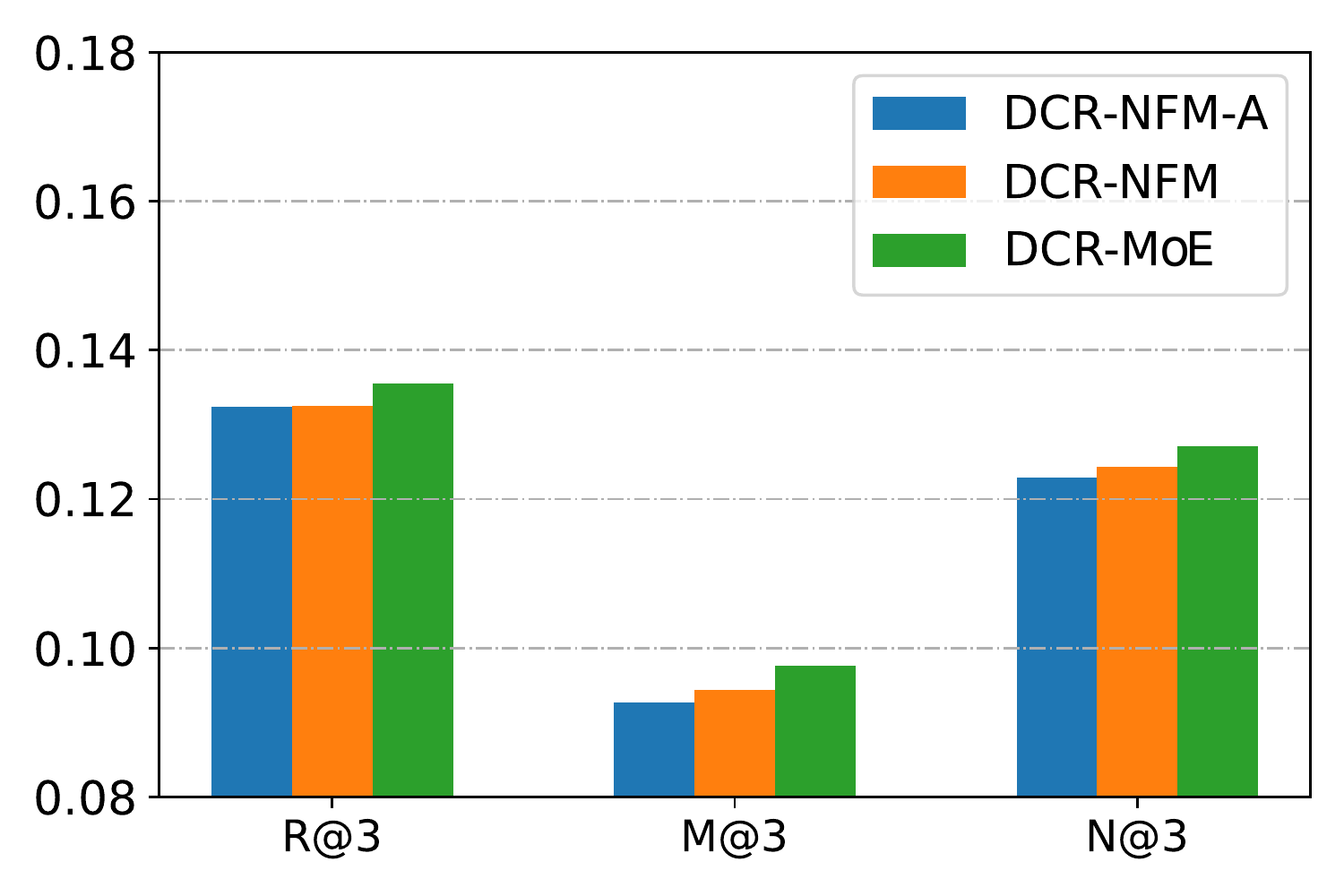}}
\caption{ Recommendation performance comparisons between different implementations for our DCR framework. \zytois{Here, R, M and N denote Recall, MAP and NDCG, respectively.}}
\label{fig:DCR-cmp}
\end{figure}

\subsubsection{The Importance of the MoE Model Architecture} Recall that one motivation for designing MoE is to speed up the inference \zy{of the DCR.} 
\zy{We compare  DCR-MoE with: 1) the DCR-NFM that implements DCR based on NFM, \ie implementing DCR without the MoE architecture, and 2) the DCR-NFM-A that speeds up the inference by NWGM approximation (Equation~\ref{eq:nwgm}), which 
can be seen as applying PD~\cite{PDA} to handle the confounding feature.
} \zy{We compare both the inference efficiency and the recommendation performance. Regarding the computation cost,} we theoretically compare the number of times to calculate the \zy{Embedding-and-Bi-interaction (EB) layers and MLPs of NFM (or MoE)} as well as \zy{report} the actual running time for inference. To fairly compare the actual running time, we run all models on the same machine with an NVIDIA RTX 3060 GPU, an Intel i7-9700K CPU, and 16 GB of memory.
Table~\ref{tab:time-cost} and Figure~\ref{fig:DCR-cmp} show the comparisons regarding time cost and recommendation performance, respectively. 
\zytoisnew{From the table and figure, we have the following observations:}

\begin{itemize}
    \item \zy{From Table~\ref{tab:time-cost}, we find: 1) DCR-MoE is sub-optimal regarding the theoretical running cost. \zytois{While DCR-MoE only requires $\frac{1}{K}$ (K=6) of DCR-NFM's EB layer calculations, DCR-MoE has more cost on running the MLPs than DCR-NFM-A.} 2) The actual running time exhibits similar trend, which justifies the ability of MoE to speed up the inference of DCR. 
Note that MoE can keep using simple experts when using more complex backbone recommender models (\eg modeling high order feature interactions). 
We thus believe that the accelerating of MoE for inference will be more significant as compared to DCR-NFM when using more complex backbone. Meanwhile, the gap to DCR-NFM-A will be reduced in such cases.} 

\item Although DCR-NFM-A achieves the best inference efficiency, its recommendation performance decreases on most metrics compared with DCR-NFM, as shown in Figure~\ref{fig:DCR-cmp}. \zytoisnew{The result can be attributed to the fact that DCR-NFM-A achieves the inference speedup by approximating computation, sacrificing the modeling fidelity.} On the contrary, DCR-MoE can also increase the recommendation performance compared with DCF-NFM, ensuring the modeling fidelity. 
\end{itemize}
\zytoisnew{The above two observations show that MoE can not only improve the efficiency of DCR but also improve its effectiveness.} \zytoisnew{Note that our DCR-MoE has the similar training time cost to the single model methods, \eg NFM-WA and DCR-NFM, since each sample is only fed into one expert and used to update one expert. We do not study the training time cost here. With the training and inference efficiency and recommendation performance considerations, we believe MoE is an important design for DCR.}

\begin{figure}
\centering
\subfigure[\textbf{Counts of samples}]{\label{fig:group_distribution} \includegraphics[width=0.49\textwidth]{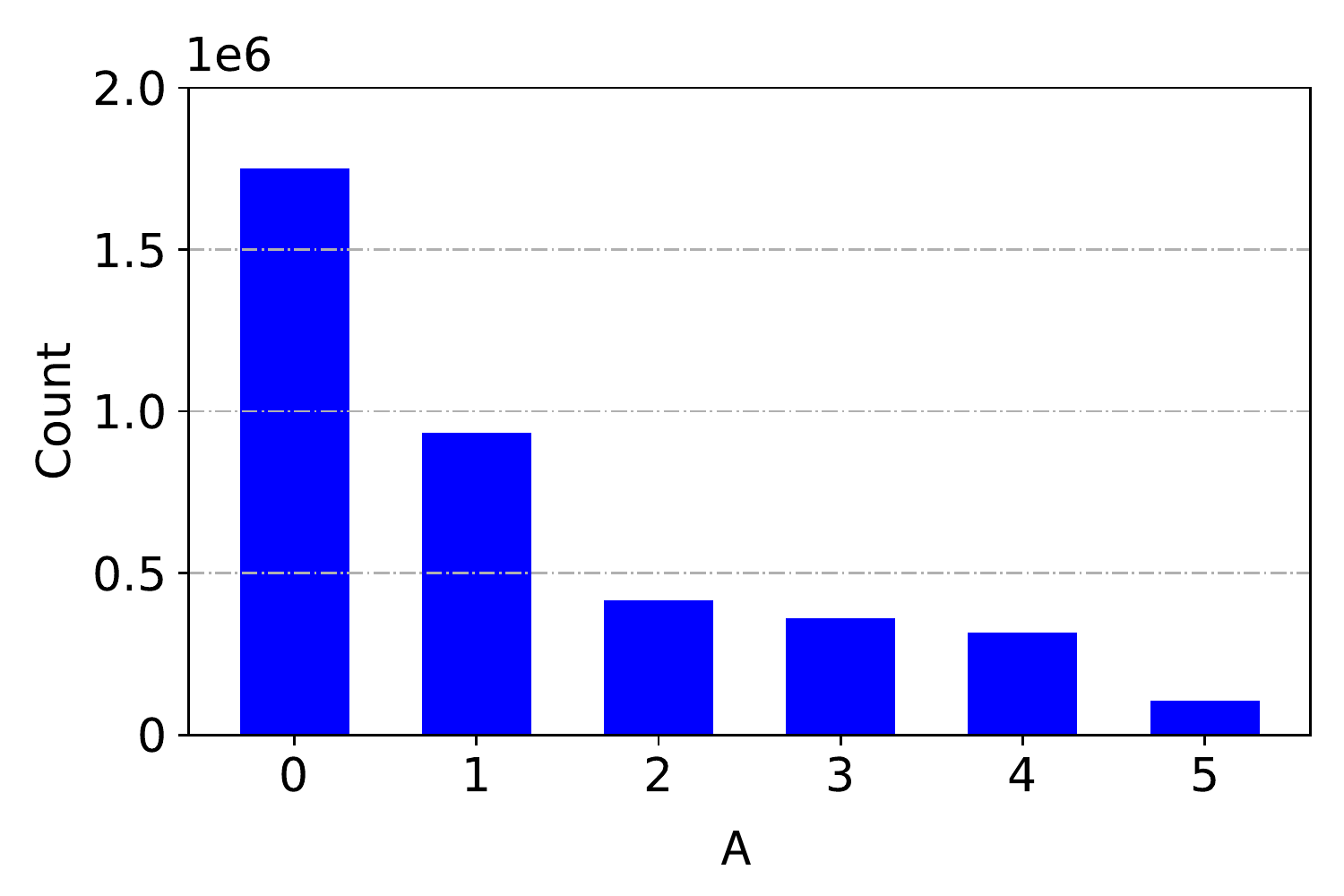}}
\subfigure[ \textbf{The difference of BCE loss}]{\label{fig:group_loss} \includegraphics[width=0.49\textwidth]{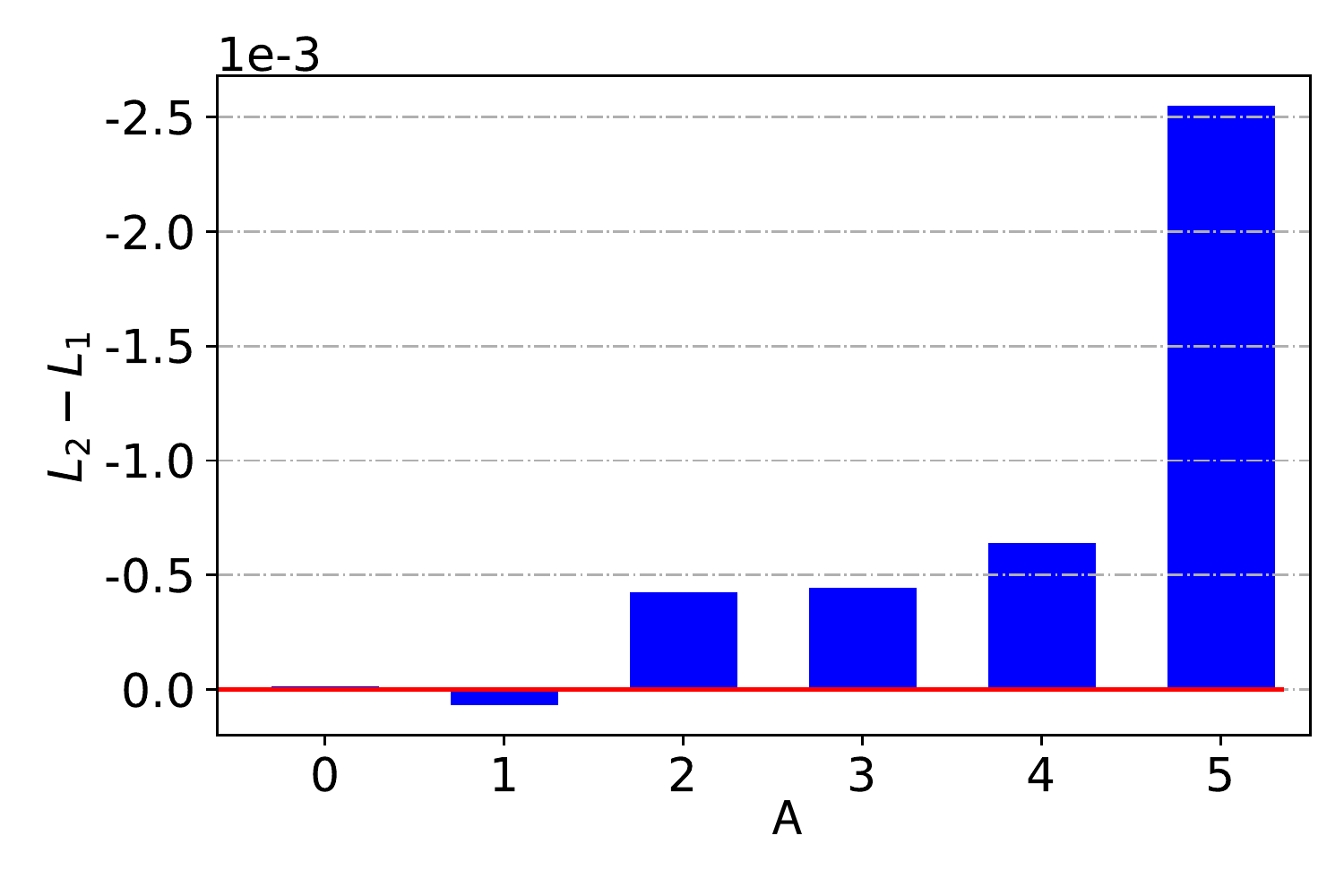}}
\caption{The advantages of MoE for fitting $P(Y|U,X,A)$ on different values of confounding feature $A$. (a) the counts of samples for different values of $A$ (sorted according to the counts). (b) The difference ($L_{2}-L_{1}$) between the BCE loss of MoE ($L_2$) and the BCE loss of NFM-WA ($L_1$) on different $A$. \zytoisnew{Please note that a higher bar in (b) means a more negative value of $L_{2} - L_{1}$.}} 
\label{fig:group-info}
\end{figure}

\subsubsection{Why Does MoE Bring Improvements?} 
\zy{We then investigate why the MoE architecture of DCR-MoE brings performance gains, especially on the dataset Kwai.} 
Note that Kwai has an extremely unbalanced distribution of $A$, as shown in Figure~\ref{fig:group_distribution}. 
\zy{We postulate that MoE enhances the estimation of $P(Y|U, X, A)$ on different values of $A$, and further, a better estimation for the $ P(Y|U,do(X))$, under an unbalanced distribution of $A$.} In a single model (NFM-WA or DCR-NFM), the model parameters are dominated by the samples with the head (\ie more frequent) values of $A$ \zytoisnew{since the training loss is dominated by them}, \zy{making samples with the tail values of $A$ cannot be well expressed. That means the learning of $P(Y|U, X, A)$ on the head values of $A$ will disturb the learning of that on the tail values.} 
However, our MoE takes different experts to represent different values of $A$, making the $P(Y|U, X, A)$ on tail values of $A$ influenced by the head values less. \zytoisnew{Thus, MoE enhances the estimation of $P(Y|U, X, A)$.} 

To verify the above intuition, we compare the \zytois{average} BCE loss of MoE (denoted as $L_2$, \zy{MoE is the one defined in Section~\ref{ab:intervention}}) and the \zytois{average} BCE  loss of NFM-WA (denoted as $L_1$) over samples with the same values of $A$. \zytoisnew{Both MoE and NFM-WA are modeling the correlation $P(Y|U,X,A)$, so the smaller BCE loss means a better estimation for $P(Y|U,X,A)$.} The results are shown in Figure~\ref{fig:group_loss}. Together with Figure~\ref{fig:group_distribution}, we can find that MoE and NFM-WA have larger loss differences on the samples with more tail values of $A$ ($A=2,3,4,5$), and the loss of NFM-WA is larger than the loss of MoE. Meanwhile, MoE has similar losses to NFM-WA on the head values ($A=0,1$). This shows the MoE model architecture can estimate $P(Y|U, X, A)$ on tail values of $A$ better and keep the accuracy of $P(Y|U, X, A)$ on other head values. \zytois{Due to the better estimation of $P(Y|U,X,A)$ on tail values of the confounding feature with MoE architecture, DCR-MoE can get a better estimation of the $P(Y|U,do(X)$, leading to the better recommendation performance.}

\begin{table}[]
\caption{{The relative improvements (RI) of DCR-MoE to baselines NFM-WA and NFM-WOA in recommendation performance, when these models (including DCR-MoE and the baselines) are performed on confounding or non-confounding features. For Kwai and Wechat, the results for the top-20 recommendation and top-5 recommendation are reported, respectively.}}
\label{tab:RI-compare}
\resizebox{0.95\textwidth}{!}{
\begin{tabular}{c|cc|cc}
\hline
Datasets             & \multicolumn{2}{c|}{Kwai(top-20)} & \multicolumn{2}{c}{Wechat(top-5)} \\ \hline
Feature type & RI to NFM-WA       & RI to NFM-WOA      & RI to NFM-WA      & RI to NFM-WOA      \\ \hline
confounding feature                  & 30.7\%          & 36.4\%          & 4.2\%          & 2.0\%           \\
non-confounding feature                   & 0.4\%           & 3.7\%          & 1.0\%          & -0.5\%          \\ \hline
\end{tabular}}
\end{table}

\subsubsection{Does DCR-MoE Really Address the Confounding Feature Issue?} We further verify that the success of DCR-MoE comes from addressing the confounding feature issue, from another perspective. Specially, we study the improvements of DCR-MoE to basic baselines (NFM-WA and NFM-WOA) in recommendation performance, forcibly performing DCR-MoE and the baselines on non-confounding features\footnote{We manually select a feature having similar correlation coefficients to the training label and the testing label as the non-confounding feature, \ie $|\rho_{1}/\rho_{2}|$ (\cf{Table~\ref{tab:data-info}}) is near to $1$. To perform models on the non-confounding feature, we forcibly replace the confounding feature $A$ with the non-confounding feature.}. If the improvements are still large, compared to the normal case where all models are performed on confounding features, the DCR-MoE may not address the confounding feature issue. Thus, we next compare the relative improvements (RI, defined in Table~\ref{tab:overall}) of DCR-MoE to the baselines in the above two cases. The results are summarized in Table~\ref{tab:RI-compare}. From the table, we find that: DCR-MoE performed on non-confounding features shows far fewer (even negative) relative improvements to the corresponding NFM-WA and NFM-WOA,  compared to the case where models are performed on confounding features. This verifies that the performance improvement of DCR-MoE (in the normal case) is brought by effectively addressing the confounding feature issue.


\begin{figure}
\centering
\subfigure[\textbf{ Kwai}]{\includegraphics[width=0.48\textwidth]{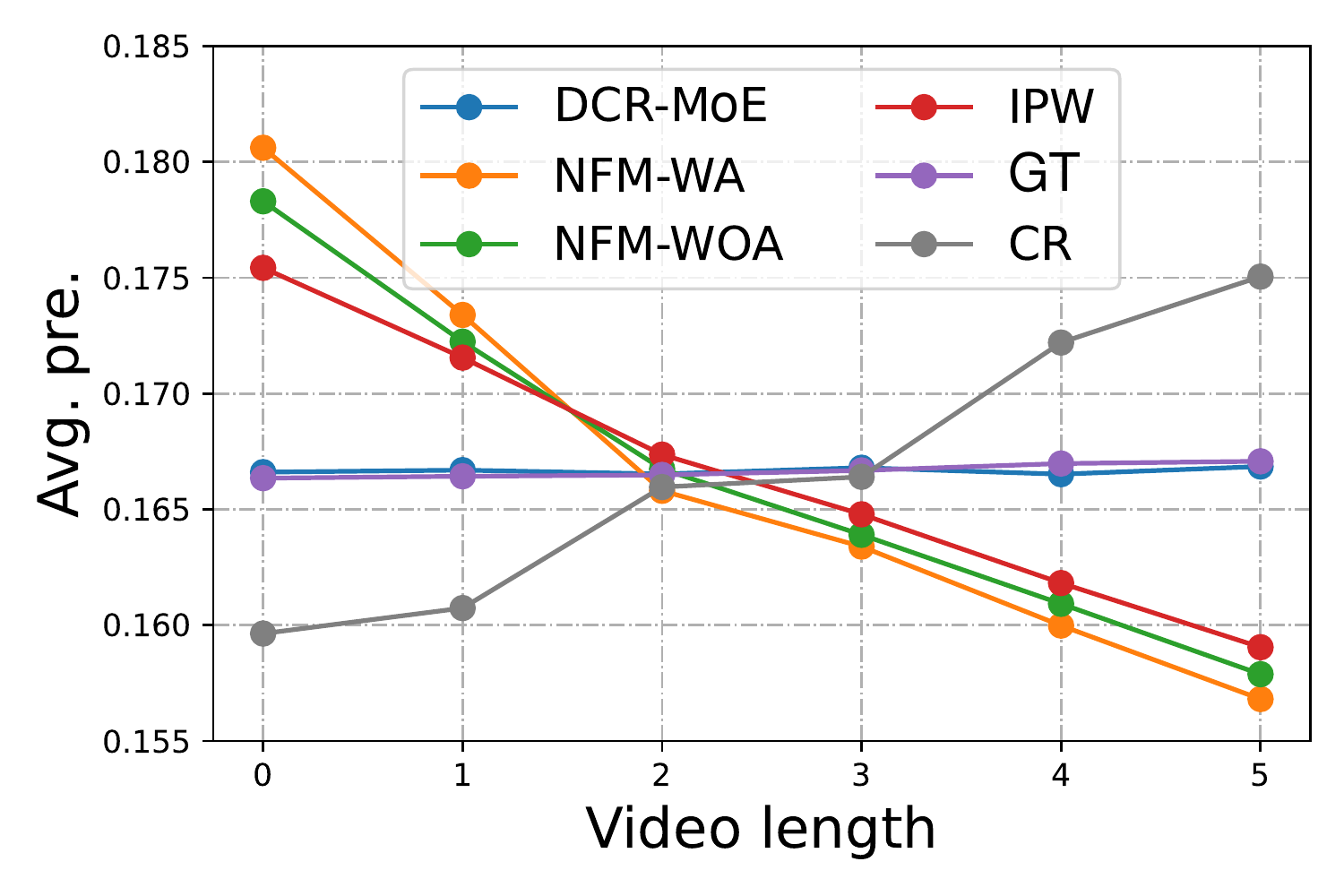}}
\subfigure[ \textbf{Wechat}]{ \includegraphics[width=0.48\textwidth]{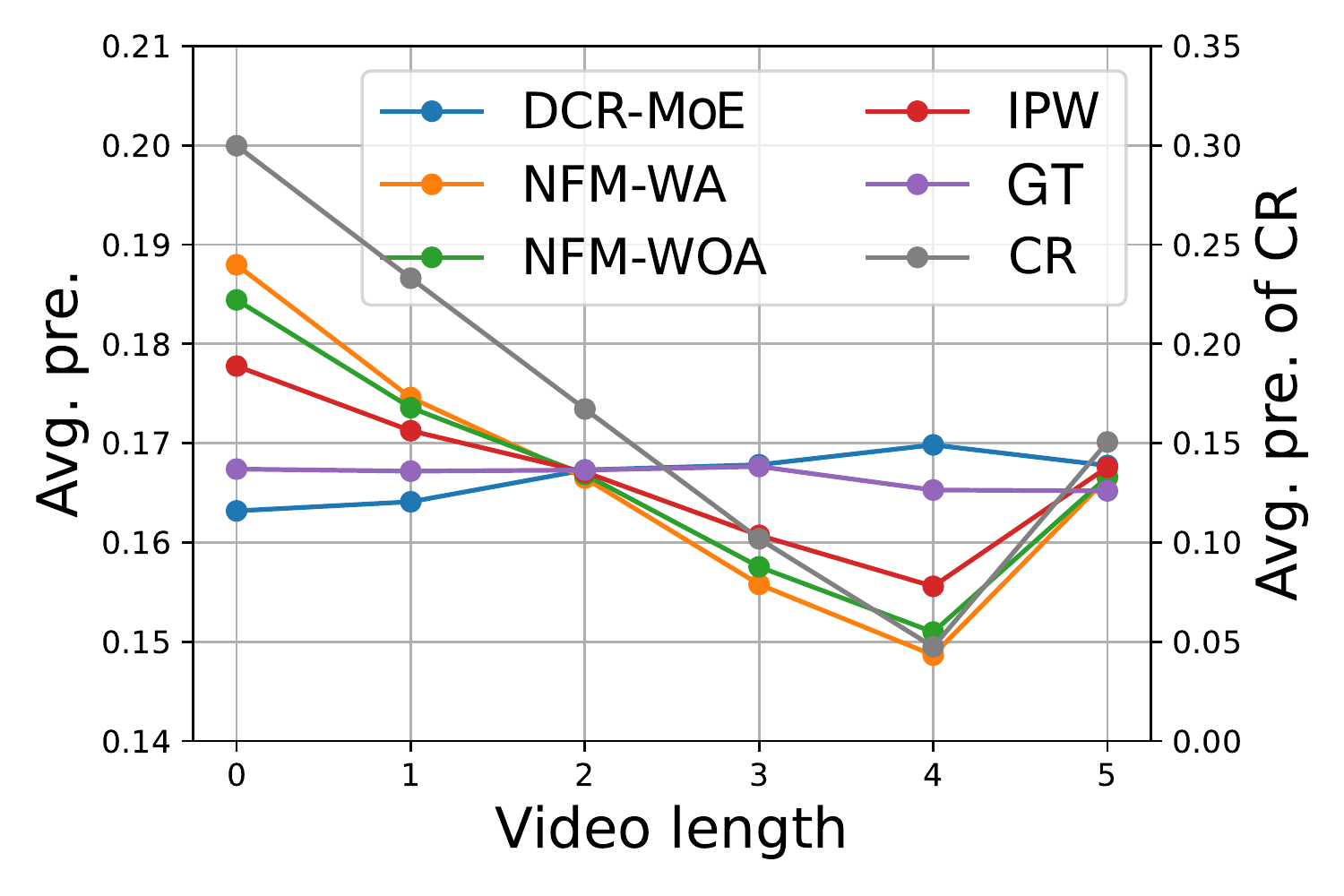}}
\caption{The average prediction (Avg. pre.) scores of different methods over samples in the same group, where the group is defined according to the value of the confounding feature, on Kwai and Wechat. \zytoisnew{"GT" is used to represent the ground-truth interest of the population to different groups.} }
\label{fig:avg-pre}
\end{figure}
\subsection{RQ3: Prediction Analyses} 

The performance study has shown the superiority of DCR-MoE regarding overall recommendation accuracy.
In this subsection, we further study whether our DCR-MoE truly \zy{eliminates the impact of the confounding feature by conducting an analysis of the model predictions.}
\zy{Specifically, we first divide items into different groups according to the values of $A$, \zytoisnew{such that items with the same value of the confounding feature belong to a group}. Then we average the prediction scores from DCR, NFM-WA, NFM-WOA, IPW, and CR over the samples that belong to the same group.} Meanwhile, we average the \zy{evaluation interaction} label (\ie testing label) over the samples \zy{that belong to the same group, which can represent the ground-truth interest of the population to different item groups. The later result is denoted as GT.} The comparison\footnote{Since the scales of average prediction scores are different for different methods. We 
adopt the softmax function to deal with the results to better reflect the shape of distributions.} between different methods is shown in Figure~\ref{fig:avg-pre}, 
where we have the following observations:




\begin{itemize}
    \item \zy{The lines of GT are almost flat, which shows the true user-item matching is less related to the confounding feature, \ie video length. Meanwhile, the lines of DCR-MoE are most similar to the lines of GT that are taken as the ground truth of user interests, meaning DCR-MoE captures the true user-item matching.} Indeed, for all user-item candidates, DCR-MoE forcibly changes their confounding feature to whatever possible values according to the same P(A) at inference, making the recommendations invariant with respect to the value of the confounding feature $A$.
    

    \item \zy{NFM-WOA and NFM-WA are both highly related to the confounding feature (since their lines are not flat). They are both biased towards some special groups, \eg giving higher scores to the group with the shortest video length. It is natural for NFM-WA since it captures the direct effect of $A$ on $Y$ for estimating user-item matching by modeling $P(Y|U,X,A)$. The biased results of NFM-WOA which models $P(Y|U,X)$ can be attributed to the spurious correlations brought by the backdoor path, and show that the impact of the confounding feature $A$ will still be captured even if $A$ is removed from the inputs.} 
    
    \item \zy{The lines of IPW are slightly flatter, compared with NFM-WA and NFM-WOA, but are still far from the lines of GT. This shows IPW can only slightly mitigate the biased recommendation, which can be attributed to the fact the propensity weights are not easy to be estimated well.}
    \item \zy{The phenomenon of CR is rather strange.} For the shortest video, CR has the lowest average prediction on Kwai, while it has the largest average prediction on Wechat. We think this is another evidence that CR cannot truly disentangle the effects of the confounding feature (undesired) and other features, leading to incorrectly removing effects.
\end{itemize}
Here we do not compare FairGo since it blindly removes all information related to the confounding feature and is straightforward to get flat lines by tuning hyper-parameters to remove more \zy{information}.

\section{Conclusion}

This work studies a new problem for dealing with the impact of item confounding feature on recommendation. 
We analyzed the problem from a causal view and recognized a backdoor path through the confounding feature, which results in spurious correlations between the remaining content features and the interaction behavior and biased recommendation. 
To remove the backdoor path, we proposed a DCR framework to estimate the causal effect of content features on the interaction, with a MoE architecture to speed up the inference. 
\zytois{We prove in theory that DCR has good generality to deal with more cases of different causal relations between confounding feature and other features.}
We conduct experiments on two real-world datasets, providing insightful analyses on the rationality and effectiveness of our proposal.

This work shows the importance of causal modeling for item features in recommendation, but focuses on one confounding feature. In future, we will extend the proposed framework in the following directions: 1) handling multiple confounding features; 2) diving into the detailed causal relations among content features; and 3) incorporating the causal relations among user features. Moreover, we would like to explore the causal discovery problem in recommendation to get rid of the labor cost on causal graph construction.

\begin{acks}
\zytois{This work is supported by the National Key Research and Development Program of China (2020AAA0\\106000) and the National Natural Science Foundation of China (U19A2079, 62121002).}
\end{acks}

\bibliographystyle{ACM-Reference-Format}
\bibliography{7-ref}


\end{document}